\documentclass{article}
\usepackage{arxiv}

\usepackage[utf8]{inputenc} 
\usepackage[T1]{fontenc}    
\usepackage{hyperref}       
\usepackage{url}            
\usepackage{booktabs}       
\usepackage{microtype}      
\usepackage{lipsum}	    	
\usepackage{graphicx}
\usepackage{doi}
\usepackage[section]{placeins}
\usepackage{mathtools}
\usepackage{amsmath}
\usepackage{amssymb}
\usepackage[ruled,vlined]{algorithm2e}
\usepackage{lineno}
\usepackage[dvipsnames]{xcolor}
\usepackage{enumitem}
\setlist[itemize]{noitemsep, topsep=0pt}

\usepackage{color}
\usepackage{soul}

\title{\textbf{I-FENN for thermoelasticity based on physics-informed temporal convolutional network (PI-TCN)}}

\author{{\hspace{1mm}Diab W. Abueidda}\thanks{da3205@nyu.edu} \\
	Civil and Urban Engineering Department\\
	New York University Abu Dhabi\\
        National Center for Supercomputing Applications\\
	University of Illinois at Urbana-Champaign\\
	\And
	{\hspace{1mm}Mostafa E. Mobasher}\thanks{mostafa.mobasher@nyu.edu}\\
	Civil and Urban Engineering Department\\
	New York University Abu Dhabi\\
}

\renewcommand{\shorttitle}{I-FENN for transient thermoelasticity}

\begin{document}

\maketitle

\begin{abstract}
Most currently available methods for modeling multiphysics, including thermoelasticity, using machine learning approaches, are focused on solving complete multiphysics problems using data-driven or physics-informed multi-layer perceptron (MLP) networks. Such models rely on incremental step-wise training of the MLPs, and lead to elevated computational expense; they also lack the rigor of existing numerical methods like the finite element method. We propose an integrated finite element neural network (I-FENN) framework to expedite the solution of coupled transient thermoelasticity. A novel physics-informed temporal convolutional network (PI-TCN) is developed and embedded within the finite element framework to leverage the fast inference of neural networks (NNs). The PI-TCN model captures some of the fields in the multiphysics problem; then, the network output is used to compute the other fields of interest using the finite element method. We establish a framework that computationally decouples the energy equation from the linear momentum equation. We first develop a PI-TCN model to predict the spatiotemporal evolution of the temperature field across the simulation time based on the energy equation and strain data. The PI-TCN model is integrated into the finite element framework, where the PI-TCN output (temperature) is used to introduce the temperature effect to the linear momentum equation. The finite element problem is solved using the implicit Euler time discretization scheme, resulting in a computational cost comparable to that of a weakly-coupled thermoelasticity problem but with the ability to solve fully-coupled problems. Finally, we demonstrate I-FENN's computational efficiency and generalization capability in thermoelasticity through several numerical examples.
\end{abstract}


\section{Introduction} \label{intro}

A multitude of nascent engineering applications presents formidable challenges to the solid mechanics community, as many of these applications require coupling two or more physical phenomena \cite{zhang2015multiphysics}, multiscale analysis \cite{weinan2011principles}, material and geometric nonlinearities, etc. For instance, thermomechanical coupling, an example of a multiphysics problem, has many applications, including metal solidification \cite{koric2010multiphysics}, geomechanics \cite{laloui2005constitutive, delage2005coupled}, biomechanics \cite{mongkol2023photo}, battery safety \cite{kato2019lithium}, additive manufacturing \cite{santi2023multiphysics, bayat2021review}, etc. Over the past few decades, the finite element method (FEM) has become crucial for comprehending and predicting intricate physical phenomena and processes \cite{hughes2012finite, wang2016fluid}. Although it is widely considered one of the most reliable and versatile numerical methods currently available, there are several challenges in the nonlinear modeling of materials and engineering processes with vast room for improving models' performance (accuracy and computation time). Recently, deep learning has been proven to be successful in various applications, and computational solid mechanics is no exception. In this paper, we propose an integrated finite element neural network (I-FENN) framework with the aim of speeding up computations by leveraging machine learning in multiphysics.

Deep learning (ML) has arisen as a tool to address some of the challenges encountered by the finite element method (FEM) in modeling nonlinear behavior. In the context of computational mechanics, deep learning models are developed using two main approaches: a data-driven approach and a physics-informed approach. Data-driven models are trained on large experimental or simulated datasets to identify complex relationships between input and output variables \cite{yan2022data, qu2021towards, heidenreich2022modeling, parrott2023multi}. By integrating ML with FEM, researchers can develop surrogate models that accurately predict the physical response to different inputs, reducing the need for expensive simulations. For example, Ghavamian et al. \cite{ghavamian2019accelerating} proposed a framework in which stress-strain data collection is streamlined in a strain-softening Perzyna viscoplasticity micro model. The data generated using the micro model are then utilized to train a surrogate model based on a recurrent neural network (RNN). The RNN-based surrogate model is then integrated into a multiscale FE scheme. Moreover, integrating the FEM and ML has also been employed to develop a concurrent optimization framework based on multiscale finite element analyses and regression neural networks \cite{zhou2022hierarchical}.

Another approach to using machine learning in computational mechanics is the physics-informed or physics-constrained technique. A physics-informed neural network (PINN) enables the incorporation of physical principles or constraints into neural networks, and it can be easily combined with the data-driven approach by devising the loss function to be minimized \cite{raissi2019physics, cuomo2022scientific}. PINNs have emerged as a promising technique to overcome the limitations of traditional data-driven models by accounting for prior knowledge of the system in the training process. By leveraging the knowledge provided by physical laws, PINNs can enhance the generalization and accuracy of the model. PINNs are used to solve various mechanics problems, including, but not limited to, elasticity \cite{guo2022energy}, plasticity \cite{he2023deep, niu2023modeling}, heat transfer \cite{cai2021physics}, fluid mechanics \cite{cai2021physics}, multiphysics \cite{niaki2021physics}, and topology optimization \cite{he2022deep}. Additionally, Masi et al. \cite{masi2021thermodynamics} proposed a framework based on neural networks (NNs), called thermodynamics-based artificial neural networks (TANN), that employs the principles of thermodynamics to model strain rate-independent processes at the material point level. The TANN approach displayed robust performance, making it a promising candidate for replacing constitutive calculations in finite element incremental formulations. Cuomo et al. \cite{cuomo2022scientific} provide a recent review of PINNs. 

One way to use PINNs is to define the loss function as the residual of partial differential equations (PDEs) at specific collocation points within the physical domain and its corresponding boundary and initial conditions. This method is commonly known as the deep collocation method (DCM) and has been employed by researchers such as Niaki et al. \cite{niaki2021physics}, Henkes et al. \cite{henkes2022physics}, and Rao et al. \cite{rao2021physics}. Another method, called the deep energy method (DEM), has been presented by Samaniego et al. \cite{samaniego2020energy} and Nguyen-Thanh \cite{nguyen2020deep}, which uses a loss function based on potential energy to solve problems involving elasticity, hyperelasticity, and piezoelectricity. This approach reduces the requirement for the differentiability of the basis function and automatically satisfies traction-free boundary conditions. Fuhg et al. \cite{fuhg2022mixed} demonstrated that the deep collocation method (DCM) and deep energy method (DEM) could not be effective in resolving fine displacement and stress features. In an attempt to address this issue, they proposed a mixed deep energy method (mDEM) where stress is an added output of the neural network, and both the strong form and potential energy are used to define the loss function.

Despite the apparent success of PINNs in solving various boundary value problems, there needs to be more understanding of the behavior of these neural networks during the optimization process, as well as why they may fail to train at times. One aspect of the challenges encountered during the optimization is the non-convexity, saddle points, and rough and uneven loss landscapes \cite{basir2023investigating}. In this paper, the loss landscapes of the trained models are shown and discussed. Moreover, training PINNs can be slow and difficult, requiring a significant effort to design the network architecture and adjust its hyperparameters, often through trial and error or grid search methods. Grid search may become impractical when dealing with a large number of hyperparameters. Additionally, training of PINNs can be complicated due to their multi-term loss function. Wang et al. \cite{wang2021understanding} investigated the gradient distribution of each term in the loss function and suggested using a gradient scaling approach to determine the weights of the loss terms. Another approach for weight specification of the loss terms was proposed by Wang et al. \cite{wang2022and}, which is based on the eigenvalues of the neural tangent kernel (NTK) matrix. 

Additionally, several variations of PINN models have been proposed in an attempt to improve the performance of PINN models, such as h-PINNs \cite{lu2021physics}, VPINNs \cite{kharazmi2019variational}, fPINNs \cite{pang2019fpinns}, and B-PINNs \cite{yang2021b}. The majority of PINN models are based on MLP architectures. Although sequence-to-sequence (Seq2Seq) learning models have been developed as data-driven models, little has been done in the context of physics-informed neural network models. Data-driven Seq2Seq models that are based on recurrent neural networks (RNN) (e.g., long short-term memory (LSTM) and gated recurrent unit (GRU)) \cite{dorbane2022exploring, yu2022elastoplastic} and temporal convolutional networks (TCN) \cite{perumal2023temporal} have shown success in capturing scenarios requiring sequential input and output data. 

Models based on TCN have several advantages over those based on RNN, especially when handling long-term dependencies. TCN models do not have recurrent connections, which makes them structurally different from RNNs and better suited for processing long sequences. This non-recurrent architecture ensures that gradients propagate consistently across all time steps, preventing issues related to vanishing or exploding gradients that can be more common in RNNs \cite{bai2018empirical, aksan2019stcn}. As a result, TCNs can learn long-term dependencies faster, and their reduced network complexity enhances this efficiency. Moreover, the parallelism inherent in convolutional layers makes it easy to parallelize TCNs on multi-core CPU or GPU computational architectures \cite{bai2018empirical}. In contrast, RNNs can have difficulties with gradient flow over long sequences, leading to unstable training. Bai et al. \cite{bai2018empirical} and Alla and Adari \cite{alla2019beginning} provide additional information on the specific architecture of TCN and the benefits it offers over RNNs.

Coupling neural networks and the finite element method intrigued several researchers. For instance, Bishara et al. \cite{bishara2023machine} developed an approach that employs a machine learning model as a complementary tool to improve the homogenization process in solving a finite element problem that involves microscale-based constitutive laws. Additionally, Yadav et al. \cite{yadav2021spde} argued in favor of utilizing deep learning to predict the stabilization parameter in the streamline upwind Petrov-Galerkin (SUPG) method and combining the neural network with the capability of the finite element method to solve partial differential equations that are singularly perturbed. Another class of methods that couples neural networks and the finite element method is the hierarchical deep-learning neural networks (HiDeNN) framework, wherein the weights and biases of deep neural networks are derived from the spatial discretization or element mesh associated with the finite element method. Consequently, the spatial discretization inherent in HiDeNN is capable of emulating the finite element approach \cite{zhang2021hierarchical, lu2023convolution, park2023convolution}.

Pantidis and Mobasher \cite{pantidis2023integrated} developed and implemented a framework, integrated finite element neural network (I-FENN), to study non-local continuum damage mechanics by integrating PINNs and FEM. Particularly, they developed a physics-informed neural network to predict the non-local equivalent strain at each material point. After that, the PINN is incorporated into the element stiffness definition to conduct the local-to-non-local strain transformation. The PINN outputs are utilized to determine the element residual vector and Jacobian matrix. In a follow-up paper, Pantidis et al. \cite{PantidisError2023} demonstrated how the error in PINN converges with the size of the network and training dataset. Furthermore, their study revealed consistent performance patterns determined by the network's topology. These findings may pave the way for improving the resilience of I-FENN and expanding the possibilities for using PINNs in engineering-related tasks.

The contributions embedded in the present study outline a pathway toward enhanced efficiency and scalability in solving thermoelastic problems and potentially other multiphysics problems. Specifically, this paper contributes to the evolving field of the confluence of machine learning and computational solid mechanics as follows:
\begin{itemize}[noitemsep, topsep=0pt]
    \item We Introduce the physics-informed temporal convolution neural network (PI-TCN) model for solving the energy equation. The model's introduction, design, implementation, and testing are discussed in this study. To the best of our knowledge, this is the first TCN to be used to develop physics-informed neural network models. Such models can be used for other time- and/or path-dependent phenomena. Additionally, an MLP-based PINN model is developed for one of the numerical examples discussed in the paper for comparison purposes, and we show that the PI-TCN model outperforms the MLP-based PINN model. In this paper, we refer to the MLP-based PINN as PINN for short.

    \item The PI-TCN model is integrated into a finite element framework to improve its efficiency and scalability. This is achieved through an integrated finite element neural network (I-FENN) framework following the general approach that was first proposed in \cite{pantidis2023integrated, PantidisError2023}. It is important to note that these papers focused on fully connected MLP networks that predicted non-local damage at singular load levels. However, this manuscript diverges from that approach in several ways. Firstly, it concentrates on fully coupled thermoelasticity, a first-order problem in time. This problem is inherently more complex than the diffusion operator in the non-local damage model. Secondly, the energy equation requires the imposition of Dirichlet boundary conditions, which raises a challenge not encountered in earlier research about non-local damage models. Finally, the PI-TCN is introduced as a single network that can handle sequential (temporal) inputs and outputs. This allows for the prediction of the response at all time steps. The resulting model solves a transient coupled thermoelasticity model at the cost of an uncoupled problem.
    
    \item In an effort to understand the loss landscape and the PI-TCN training efficiency, we adapt the technique first presented by Li et al. \cite{li2018visualizing} to visualize the loss function in multiple dimensions. The visualizations lead to critical insights into the network training and open the door to new research venues.
\end{itemize}

The structure of the paper is outlined as follows: Section \ref{ProblemStatement} describes the principles behind the equations governing the thermoelasticity problem as well as the corresponding weak form and establishes the motivation for the proposed I-FENN model. Section \ref{deeplearning} discusses the deep learning techniques used in this paper and introduces the general setup for the PINN and PI-TCN models as well as the random Fourier feature mapping and loss landscape visualization methods. Section \ref{I-FENN} introduces the I-FENN framework for thermoelasticity and discusses the setup for a general problem. In Section \ref{results}, a few thermoelasticity examples are solved using the I-FENN framework. The paper concludes in Section \ref{conclu} by summarizing the key results and indicating potential directions for future research.
\section{Problem statement and motivation}\label{ProblemStatement}

\subsection{Thermoelasticity} \label{thermoelast}

\begin{figure}[!htb]
    \centering
    \includegraphics[width=0.4\textwidth]{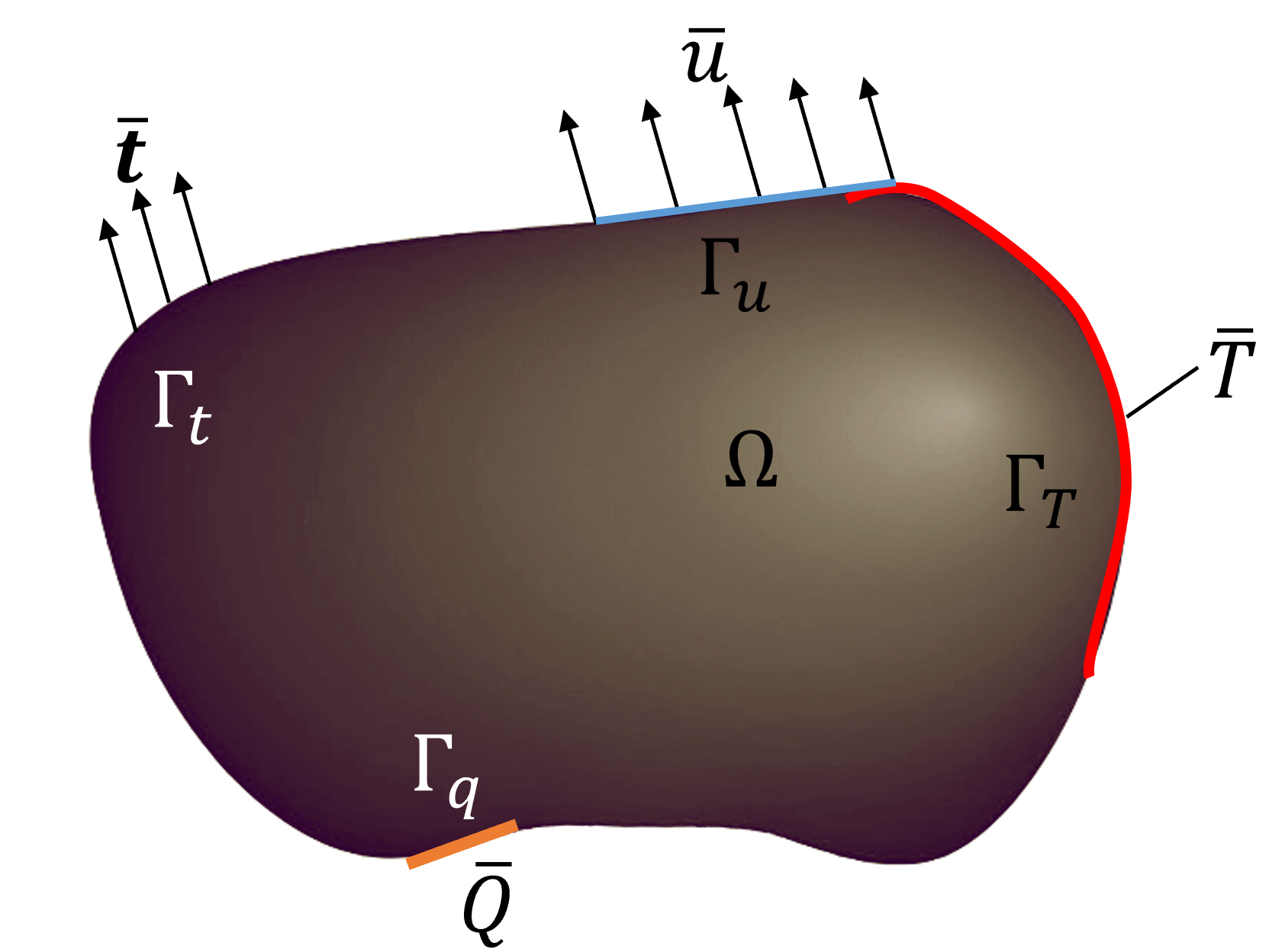}
    \caption{Schematic representation of a body with prescribed displacement, temperature, traction, and flux boundary conditions.}
    \label{potato}
\end{figure}

A frequently encountered multiphysics problem is thermoelasticity, which involves thermal strains and connects heat conduction in solids with mechanical deformation \cite{verhas2008thermoelasticity}. In this paper, we address a problem of transient thermoelastic evolution, where the thermo-mechanical fields are fully coupled. Nevertheless, we consider the evolution to be quasi-static, i.e., inertial effects are neglected. An object with isotropic and homogeneous thermoelastic properties under small deformation is considered, where the heat source and body force are assumed to be zero. The linearized thermoelasticity is regarded around a reference temperature $T_{o}$, where the system of differential equations characterizing the problem is expressed as \cite{gurtin1973linear, abeyaratne1998continuum, truesdell2013linear}:
\begin{subequations}\label{PDEs}
\begin{align}
    \boldsymbol{\nabla}\cdot \boldsymbol{\sigma}&=\boldsymbol{0}{,} \quad  \boldsymbol{x}\in\Omega, \label{BLM_eqn} \\
    \rho T_{o} \dot{\eta} + \boldsymbol{\nabla}\cdot \boldsymbol{q} &= 0, \quad  \boldsymbol{x}\in\Omega, \label{Energy_eqn}
\end{align}
\end{subequations}
where $\boldsymbol{\sigma}$ denotes the Cauchy stress tensor, $\boldsymbol{q}$ is the heat flux vector, $\eta$ marks the entropy per unit mass, and $\Omega$ is the physical domain (see Figure \ref{potato}). $\rho$ is the material density. $\boldsymbol{\nabla}$ indicates the gradient operator, and $\boldsymbol{\nabla}\cdot$ is the divergence operator. The object is subject to a set of essential and natural boundary conditions:
\begin{equation}
\begin{aligned}
    \boldsymbol{u}&=\overline{\boldsymbol{u}}, \quad \! \boldsymbol{x}\in\Gamma_{u},\\
    \boldsymbol{t}&=\overline{\boldsymbol{t}}, \quad  \boldsymbol{x}\in\Gamma_{t},\\
    T&=\overline{T}, \quad \! \boldsymbol{x}\in\Gamma_{T},\\
    Q&=\overline{Q}, \quad  \boldsymbol{x}\in\Gamma_{q}
\end{aligned}
\end{equation}
where the different boundary conditions are applied on the boundary $\Gamma=\Gamma_t \cup \Gamma_u=\Gamma_T \cup \Gamma_q$. The traction is defined as $\boldsymbol{t} = \boldsymbol{\sigma}\cdot\boldsymbol{n}$ on the traction boundary part $\Gamma_t$, where $\boldsymbol{n}$ represents the normal unit vector. The flux boundary condition is defined as $Q=\boldsymbol{q}\cdot\boldsymbol{n}$ on the flux boundary part $\Gamma_q$. Temperature and displacement boundary conditions are imposed on the temperature boundary part $\Gamma_T$ and displacement boundary part $\Gamma_u$, respectively. 

Equations \ref{BLM_eqn} and \ref{Energy_eqn} represent the balance of linear momentum and energy equation, respectively. The heat flux vector $\boldsymbol{q}$ is given by Fourier's law \cite{gurtin1973linear, abeyaratne1998continuum, truesdell2013linear}:
\begin{equation}
\begin{aligned}
    \boldsymbol{q} = -k \boldsymbol{\nabla} T,
    \label{FL}
\end{aligned}
\end{equation}
where $T$ is the temperature, and $k$ is the thermal conductivity of the material. $\boldsymbol{u}$ denotes the displacement vector. Since small deformation is presumed, the strain is given by:
\begin{equation}\label{eps}
\begin{aligned}
    \boldsymbol{\varepsilon}=\frac{1}{2}(\boldsymbol{\nabla} \boldsymbol{u}+\boldsymbol{\nabla} \boldsymbol{u}^T),
    \end{aligned}
\end{equation}
where $\boldsymbol{\varepsilon}$ denotes the infinitesimal strain tensor. The thermoelastic constitutive equations are expressed as \cite{gurtin1973linear, abeyaratne1998continuum, truesdell2013linear}:
\begin{subequations}\label{constitutive}
\begin{align}
    \boldsymbol{\sigma} &= 2 \mu \boldsymbol{\varepsilon} + \left(\lambda \text{tr}(\boldsymbol{\varepsilon}) - \alpha (3\lambda+2\mu) ( T - T_{o} )\right)\boldsymbol{I}, \label{sigma} \\
    \rho T_{o} \dot{\eta} &= \rho C_{\varepsilon} \dot{T} + \alpha (3\lambda+2\mu) T_{o} \text{tr}(\boldsymbol{\dot{\varepsilon}}), \label{entropy}
\end{align}
\end{subequations}
where $C_{\varepsilon}$ is the specific heat per unit mass at constant strain, $\text{tr}()$ is the trace operator, $\alpha$ is the coefficient of thermal expansion, and $\mu$ and $\lambda$ are Lamé constants. Substituting Equation \ref{entropy} in Equation \ref{Energy_eqn} yields:
\begin{equation}\label{new_energy}
\begin{aligned}
    \rho C_{\varepsilon} \dot{T} + \alpha (3\lambda+2\mu) T_{o} \text{tr}(\boldsymbol{\dot{\varepsilon}}) +  \boldsymbol{\nabla}\cdot \boldsymbol{q} &= 0.
\end{aligned}
\end{equation}

\subsection{Finite element analysis}\label{FEA}

The finite element analysis uses the open-sourced FEniCS library \cite{alnaes2015fenics}. The finite element analysis performed in this paper uses a monolithic approach as the reference solution. Using this approach, temperature and displacement fields are coupled and solved simultaneously. It is worth pointing out that a staggered approach could have been used instead \cite{farhat1991unconditionally, kristensen2020phase}. In this alternative approach, one of the variables would be calculated first using the predicted values of the other variable, followed by computing the other variable in a second step using the predicted values of the first one.

Using the stress constitutive equation (Equation \ref{sigma}) and the balance of linear momentum (Equation \ref{BLM_eqn}), the mechanical weak form reads as
\begin{equation}
\begin{aligned} \label{mech_weak}
\int_{\Omega} \left(\lambda\text{tr}(\boldsymbol{\varepsilon})\boldsymbol{I}+2\mu\boldsymbol{\varepsilon} -\alpha(3\lambda+2\mu)
(T-T_0)\boldsymbol{I}\right) :\nabla^s\widehat{\boldsymbol{w}}\text{ d} \Omega &=
\int_{\Gamma} \widehat{\boldsymbol{w}} \boldsymbol{t} d\Gamma \quad \forall \widehat{\boldsymbol{w}}\in W_u 
\end{aligned}
\end{equation}
where $W_u$ denotes the displacement function space, and $\widehat{\boldsymbol{w}}$ is the displacement test function. Combining Fourier's law (Equation \ref{FL}) and the energy equation (Equation \ref{new_energy}), the thermal weak form is expressed as:
\begin{equation}
\begin{aligned} \label{th_weak}
\int_{\Omega}\left(\rho C_{\varepsilon}\dot{T} + \alpha(3\lambda+2\mu)
T_0\text{tr}(\dot{\boldsymbol{\varepsilon}})\right) \widehat{T}d\Omega + \int_{\Omega} k \nabla T\cdot\nabla \widehat{T}d\Omega= \int_{\Gamma} k\partial_n T \widehat{T} d\Gamma \quad \forall \widehat{T} \in W_T
\end{aligned}
\end{equation}
where $\partial_n$ denotes the normal derivative, $W_T$ denotes the temperature function space, and $\widehat{T}$ is the temperature test function. An implicit Euler scheme is adopted; the thermal weak can be expressed incrementally as: 
\begin{equation}
\begin{split} \label{dis_th_weak}
\int_{\Omega}\left(\rho C_{\varepsilon}\dfrac{T_{n+1}-T_n}{\Delta t} + \alpha(3\lambda+2\mu) T_0\text{tr}\left(\dfrac{\boldsymbol{\varepsilon}_{n+1}-\boldsymbol{\varepsilon}_n}{\Delta t}\right)\right) \widehat{T}d\Omega \\ + \int_{\Omega} k \nabla T\cdot\nabla \widehat{T}d\Omega  = \int_{\Gamma} k\partial_n T \widehat{T} d\Gamma \quad \forall \widehat{T} \in W_T.\\
\end{split}
\end{equation} 

\subsection{Introduction to I-FENN}\label{I_FENN_Intro}
When the I-FENN framework is used, and the physics-informed machine learning model (PINN or PI-TCN) is trained, the finite element is not needed to solve the energy equation, as its effect is now inherited in the trained physics-informed machine learning model:
\begin{equation}
\begin{split} \label{T_network}
T\left(\boldsymbol{x}, t \right)&=\mathcal{N}\mathcal{N}\left(\boldsymbol{x}, t; \boldsymbol{\phi}^{*} \right)\\
\end{split}
\end{equation} 
where $\mathcal{N}\mathcal{N}$ refers to the trained neural network model, and $\boldsymbol{\phi}^{*}$ denotes the optimized set of weights and biases. The details of the $\mathcal{N}\mathcal{N}$ models used in this paper are discussed in detail in the subsequent sections. Table \ref{FEvsIFENN} summarizes the unknowns and weak form for the fully coupled finite element method and the FE analysis done within the I-FENN framework. The I-FENN framework computationally decouples the energy equation from the linear momentum equation, but its effect is captured via a physics-informed machine learning model (PINN or PI-TCN). Compared to recent advancements in the field, as outlined in papers \cite{haghighat2022physics, zlatic2023incompressible, harandi2023mixed, amini2022physics}, the I-FENN framework for thermoelasticity stands out as a distinct method. It combines the finite element method (FEM) with neural networks to create a reliable and efficient numerical solution. Unlike most research on multiphysics deep learning models focusing on MLP networks to replace the FE scheme or traditional solvers, the proposed I-FENN approach integrates the physics-informed temporal convolutional network into a finite element scheme. Such an approach has the potential to significantly reduce computational costs while maintaining the robustness of conventional FEM solvers.

\begin{table}[]
\centering
\caption{Comparison between the fully coupled finite element method and proposed I-FENN framework.}
\label{FEvsIFENN}
\begin{tabular}{c|cc}
\hline
& fully coupled finite element method                  & I-FENN framework          \\ \hline
FEM unknowns   & $\boldsymbol{u}$, $T$                                   & $\boldsymbol{u}$                \\
\multicolumn{1}{l|}{Weak form} & \multicolumn{1}{l}{Equation \ref{mech_weak} and Equation \ref{dis_th_weak}} & \multicolumn{1}{l}{Equation \ref{mech_weak}} \\ \hline
\end{tabular}
\end{table}
\section{Physics-informed machine learning} \label{deeplearning}

This section briefly introduces deep learning techniques and how physical laws are incorporated to solve partial differential equations governing a system of interest. This section paves the way by discussing the physics-informed machine learning models used within the I-FENN framework, as detailed in Section \ref{I-FENN}. Section \ref{MLP} summarizes the architecture of the multilayer perceptron neural network and presents a general PINN model, while Section \ref{TCN} discusses the TCN-based Seq2Seq learning model and introduces the physics-informed temporal convolutional network (PI-TCN). Section \ref{RFF} discusses the Fourier feature mapping used in the developed PINN and PI-TCN models, as discussed later. Section \ref{Loss_vis} talks about the visualization of the loss landscapes of trained models.

\subsection{Multilayer perceptron PINNs}\label{MLP}

\begin{figure}[!htb]
 \centering
    \includegraphics[width=0.65\textwidth]{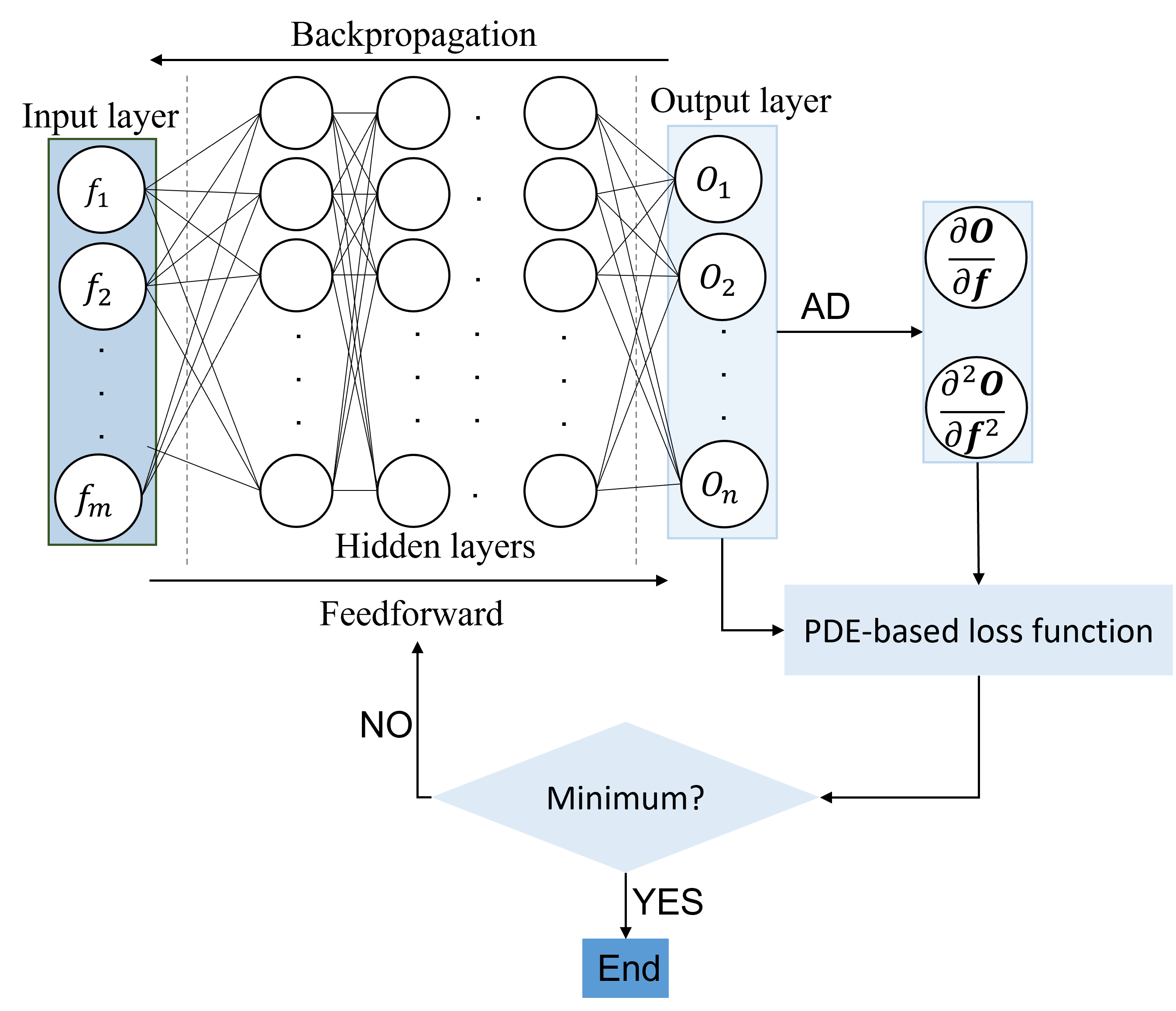}
    \caption{Physics-informed neural network based on MLP neural networks.}
    \label{mlp_fig}
\end{figure}

Multilayer perceptron (MLP) neural networks (fully connected) consist of interconnected neurons arranged in layers. The neurons are organized into linked layers, transmitting information from one layer to the subsequent layer. The network is designed to learn through a backpropagation mechanism, with the depth of the network being determined by the number of hidden layers between the input and output layers. Figure \ref{mlp_fig} illustrates a fully connected neural network. For a layer $l$, the output $\boldsymbol{\hat{Y}}^{l}$ is written as:
\begin{equation}\label{mlp_eqn}
\begin{aligned}
    \boldsymbol{Z}^{l}&=\boldsymbol{W}^{l} \boldsymbol{\hat{Y}}^{l-1}+\boldsymbol{b}^{l}\\
    \boldsymbol{\hat{Y}}^{l}&=f^{l}\left(\boldsymbol{Z}^{l}\right)\\
\end{aligned}
\end{equation}
where $\boldsymbol{W}$ and $\boldsymbol{b}$ denote weights and biases, respectively. The activation function $f^{l}$ is an $\mathbb{R} \rightarrow \mathbb{R}$ mapping that transforms vector $\boldsymbol{Z}^{l}$ into the output of the corresponding layer. Backpropagation is used throughout the training process, and the loss function, $\mathcal{L}$, is iteratively minimized to update the weights and biases.

Figure \ref{mlp_fig} illustrates a generic PINN model where $\boldsymbol{f}$ is the input features, $m$ is the number of input features, $\boldsymbol{O}$ is the output vector, and $n$ is the length of the output vector. Typical input features for a PINN model are the coordinates. Starting with initial weights and biases for the PINN model, the input features are fed in consecutive hidden layers, and the output vector is obtained. Then, using automatic differentiation (AD), the first- and second-order derivatives are computed, assuming the PDE governing this arbitrary system is only second-order. The loss function is then computed based on the residual of the PDE. The loss function is minimized to update the weights and biases.

\subsection{Physics-informed temporal convolutional network (PI-TCN)}\label{TCN}

Temporal Convolutional Networks (TCN) are a type of convolutional neural network (CNN) architecture \cite{bai2018empirical}. CNNs are a commonly used approach for analyzing images and obtaining insights from them. The TCN incorporates Conv1D layers, which are one-dimensional convolutional layers. Convolution involves applying a smaller matrix, also known as a kernel or filter, to a larger matrix representing a domain. During each application, the corresponding elements of the two matrices are multiplied and combined together. The kernel is then moved to a different section of the domain matrix, and a feature map is produced for a given kernel. The backpropagation process calculates the gradients of the loss function with respect to the kernel weights, and the weights are adjusted in a way similar to fully connected neural networks.

\begin{figure}[!htb]
    \centering
    \includegraphics[width=1.0\textwidth]{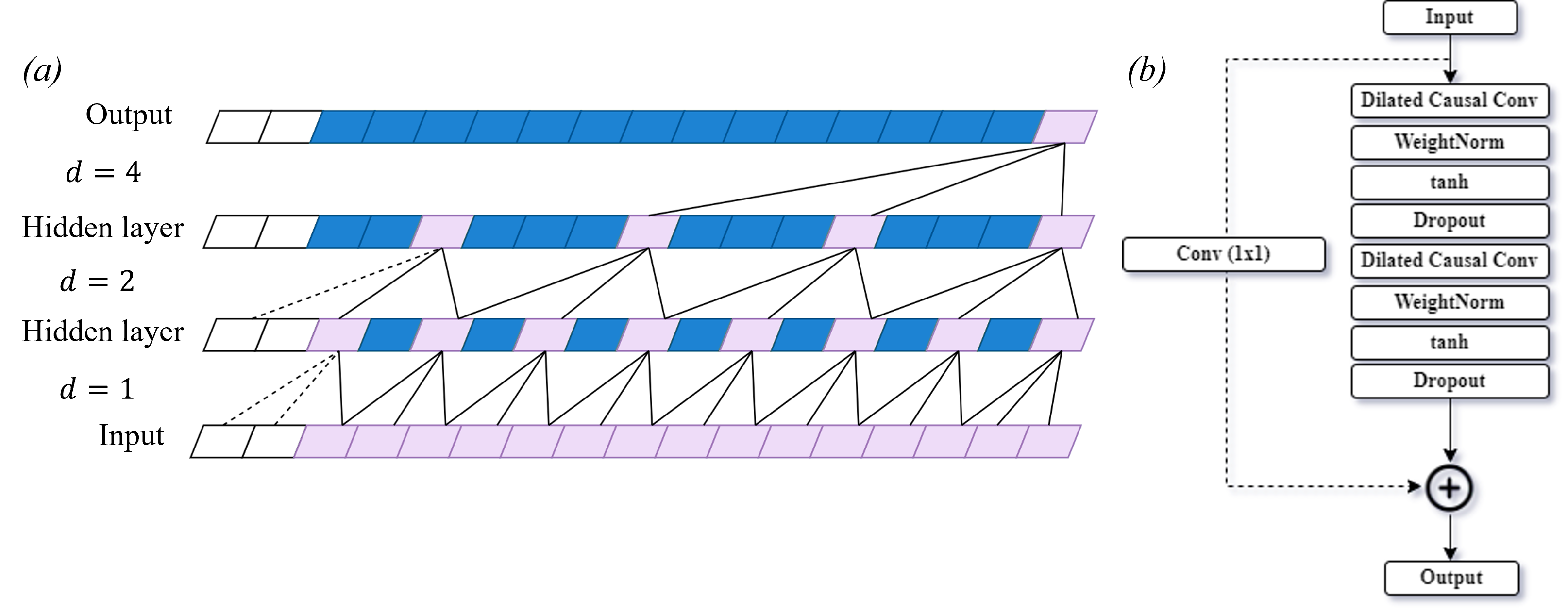}
    \caption{ Illustration for the TCN architecture: (a) A stack of dilated convolutions with dilation factors $d = 1,2,4$ and a filter size $k_{size} = 3$. The pink elements in the sequences are the ones whose information was propagated to a specific pink element in the output sequence. For example, the last element in the output sequence can access information from all the elements from the input sequence and other pink elements in the hidden layers.} (b) TCN residual block.
    \label{tcn_fig}
\end{figure}

Figure \ref{tcn_fig} shows the TCN architecture. The TCN residual block has two paths: a main path and a residual connection, as shown in Figure \ref{tcn_fig}b. The residual connection is shown in dotted lines, whereas the main path is shown in solid lines. The main path has several layers: dilated causal convolution, dropout, weight normalization, and activation function. The dilated convolution operation (shown in Figure \ref{tcn_fig}a) is expressed as:
\begin{equation}\label{dilated_conv}
\begin{aligned}
 C_d\left(s\right) &= \sum_{i=0}^{k_{size}-1} F\left( i\right) \cdot S_{s-d \cdot i}
\end{aligned}
\end{equation}
where $C_d\left(s\right)$ is the output of the dilated convolution operation at position $s$ of a sequence, and $F\left( i\right)$ represents the filter or kernel values at each position $i$. $S$ is the input sequence, and the position is defined by the subscript $s-d \cdot i$. $s$ is the current position in the output sequence, $d$ is the dilation rate, and $i$ is the current position in the kernel. $d \cdot i$ is the number of steps skipped from the current position $s$; this skipping of steps is what creates the "dilated" effect. Dropout \cite{srivastava2014dropout} and weight normalization (WeightNorm) \cite{salimans2016weight}, shown in Figure \ref{TCN}b, are typical techniques in deep learning to enhance the performance of neural networks \cite{chollet2021deep}. Residual connections let information bypass the layer transformation process (main path) and go directly to the output, simplifying training and preventing vanishing gradients \cite{he2016deep}. For the TCN architecture used here, the residual connection consists of a 1x1 convolution ($1\times1$ Conv), which is essential in adjusting data dimensionality. In other words, it ensures that the number of filters in the residual data (output of the residual connection) matches the main data (output of the main path), allowing for element-wise addition and accounting for discrepant input-output widths \cite{bai2018empirical}.

Since the energy equation residual contains second-order derivatives, it is necessary to use activation functions at least twice differentiable. Otherwise, these derivatives will trivially reduce to zero. Thus, the hyperbolic tangent (tanh) is used as an activation function, as illustrated in the TCN residual block (see Figure \ref{tcn_fig}b). A receptive field (RF) of the TCN is defined as the number of past input time steps the TCN views when predicting the output at a particular time step. Based on the dilation, the receptive field $r_{field}$ reads as:
\begin{equation}\label{receptive}
\begin{aligned}
r_{field}&=1+2\left(k_{size}-1\right) N_{stack} \sum_{l}^{L}\ d_l
\end{aligned}
\end{equation}
where  $N_{stack}$ denotes the number of stacks, $d_l$ is the dilation rate for the $l^{th}$ layer, $L$ is the number of layers per stack, and $k_{size}$ is the kernel size of the convolutional filter. Additional information regarding utilizing the TCN network structure for detecting anomalies and its benefits compared to the recurrent neural networks can be located in the work of Bai et al. \cite{bai2018empirical} and Alla et al. \cite{alla2019beginning}.

\begin{figure}[!htb]
    \centering
    \includegraphics[width=0.9\textwidth]{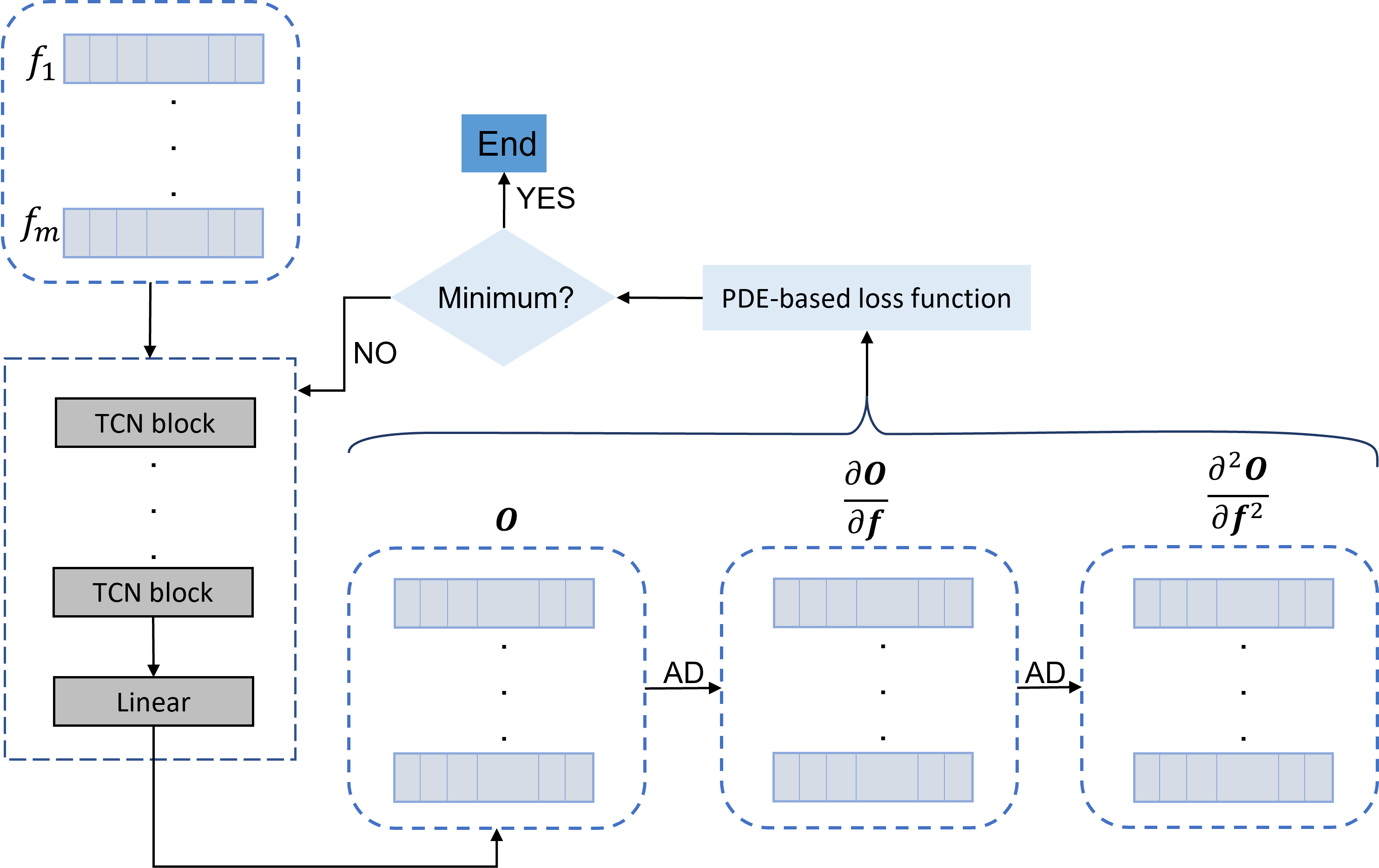}
    \caption{Proposed physics-informed temporal convolutional network (PI-TCN).}
    \label{generic_pitcn}
\end{figure}

In this paper, we propose a physics-informed version of the TCN setup, which is illustrated in Figure \ref{generic_pitcn} for a generic physics-based model. The sequential input features $\boldsymbol{f}$ are passed through one or more TCN residual blocks, where the architecture of the TCN residual block is shown in Figure \ref{tcn_fig}b. The output of the TCN residual blocks is fed into a linear (fully connected) layer, which gives the sequential outputs (solution). The first- and second-order derivatives are computed using automatic differentiation. Note that, as depicted in Figure \ref{generic_pitcn}, the output vector and its derivatives are sequential. These variables are then used to evaluate the physics-based loss function, which is iteratively minimized to attain the optimized weights and biases.

\subsection{Random Fourier feature (RFF) mapping}\label{RFF}

Tancik et al. \cite{tancik2020fourier} demonstrated that a standard multilayer perceptron (MLP) fails to learn high frequencies in theory and practice based on the neural tangent kernel (NTK) theory. They showed that a simple Fourier feature mapping could help an MLP learn high-frequency functions in low-dimensional problem domains. Although the use of Fourier feature mapping is unfamiliar with the development of PINN models, a few attempts have proven the effectiveness of using such mapping in developing a PINN model. Wang et al. \cite{wang2021eigenvector} discussed the limitation of PINNs via the neural tangent kernel theory and illustrated how PINNs are biased toward learning functions along the dominant eigen directions of their limiting NTK. Additionally, they have shown that using Fourier feature mappings with PINNs can modulate the frequency of the NTK eigenvectors. Chadha et al. \cite{chadha2022optimizing} investigated the effect of the random Fourier feature (RFF) mapping on the elasticity problem. In their hyperparameters optimization framework, they compared the results with and without RFF mapping. They concluded that introducing RFF mapping enhanced the accuracy of the PINN model and could facilitate the transferability of a set of hyperparameters optimized for a specific problem to more comprehensive geometries and boundary conditions. 

Here, a Fourier feature mapping $\gamma$ is used, where the function $\gamma$ maps the outputs of the previous layer to the surface of a higher-dimensional hypersphere using a set of sinusoids:
\begin{equation}\label{Fourier}
\begin{aligned}
    \gamma\left(\boldsymbol{X}\right) &= \left[ a_1 cos\left(2\pi\boldsymbol{h}_{1}^{T} \boldsymbol{X} \right), a_1 sin\left(2\pi\boldsymbol{h}_{1}^{T} \boldsymbol{X} \right),...,a_m cos\left(2\pi\boldsymbol{h}_{m}^{T} \boldsymbol{X} \right), a_m sin\left(2\pi\boldsymbol{h}_{m}^{T} \boldsymbol{X} \right)\right]\\
\end{aligned}
\end{equation}
where $\boldsymbol{h}_{i}$ represent the Fourier basis frequencies used for approximation, and $a_{i}$ are the corresponding Fourier series coefficients. Gaussian random Fourier feature (RFF) mapping is used. Each entry in $\boldsymbol{h}$ is sampled from a normal distribution $\mathcal{N}\left(0,\sigma^2\right)$, where $\sigma^2$ denotes the variance of the distribution. In other words, $\sigma$ is the standard deviation of the distribution $\mathcal{N}$, which is a hyperparameter that has to be specified for each problem. $\boldsymbol{X}$is the output vector from the previous layer, which is the input to the mapping function $\gamma$.

\subsection{Visualizing the loss landscapes}\label{Loss_vis}

Visualizing the loss landscape offers an understanding of the optimization problem, revealing features such as saddle points, the number of local minima, and the contours of the valleys signifying potential solutions. This knowledge assists in comprehending why specific optimization techniques achieve satisfactory outcomes, whereas others become trapped in poor local minima. Li et al. \cite{li2018visualizing} showed that more than simple visualization techniques might be needed to effectively represent the local geometry of loss function minimizers to avoid inadequate comprehension of the optimization issue. To tackle this problem, they suggested a filter normalization method that better depicts the relationship between the minima's sharpness and the model's generalization error. We must be careful when interpreting reduced-dimensional plots, as non-convexity in such a plot reveals the existence of non-convexity in the full-dimensional surface. Nonetheless, observing convexity in a low-dimensional representation does not necessarily imply that the high-dimensional function is convex \cite{li2018visualizing}.

In this paper, we utilize the filter normalization technique proposed by Li et al. \cite{li2018visualizing} to visualize the loss landscape of the trained PI-TCN model. A center $\boldsymbol{\phi}^{*}$ is chosen, and two filter-wise normalized random (sampled from a Gaussian distribution) vectors $\boldsymbol{\xi}$ and $\boldsymbol{\psi}$ are defined. This helps us visualize the shape of the loss landscape around the optimized weights. Then, a function of the following form is plotted using a log scale:
\begin{equation}\label{loss_landscape}
\begin{aligned}
    f \left(\epsilon_1, \epsilon_2\right)&=\mathcal{L} \left(\boldsymbol{\phi}^{*} + \epsilon_1 \boldsymbol{\xi} + \epsilon_2 \boldsymbol{\psi} \right),\\   
\end{aligned}
\end{equation}
where $\boldsymbol{\phi}^{*}$ is the state of the trained PI-TCN model, i.e., the optimized weights and biases. $\epsilon_1 \in \left[-1,1\right]$ and $\epsilon_2 \in \left[-1,1\right]$ are steps in the direction of the random vectors $\boldsymbol{\xi}$ and $\boldsymbol{\psi}$, respectively. Please note that the technique of filter normalization for loss landscape visualization is general and not limited to PI-TCN.
\section{Integrated finite element neural network (I-FENN) framework}\label{I-FENN}

This paper attempts to integrate neural networks into the finite element scheme to solve multiphysics problems. Specifically, the thermoelasticity problem is considered, where we employ deep learning to approximate the temperature profiles. The energy equation and the corresponding boundary conditions are cast into the minimization problem in the context of machine learning. Thus, a definition for the loss function is required, and then the loss function can be minimized within a deep learning framework such as PyTorch \cite{NEURIPS2019_9015}.

\subsection{I-FENN for thermoelasticity: Overview}\label{Section_ifenn_overview}

\begin{figure}[!htb]
    \centering
    \includegraphics[width=0.9\textwidth]{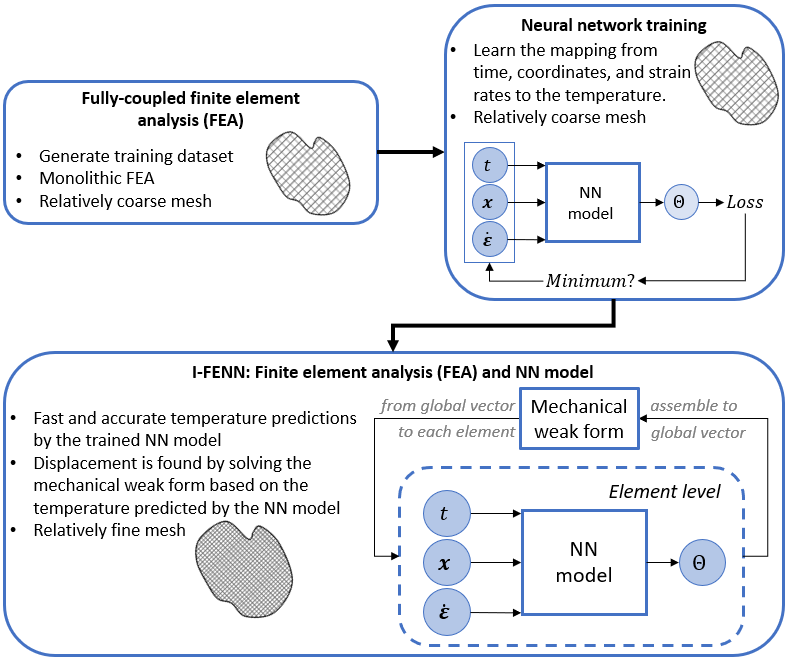}
    \caption{Illustration of the I-FENN framework for thermoelasticity. The pre-trained NN is embedded inside the weak form of the balance of linear momentum of the thermoelasticity problem.}
    \label{I-FENN_fig}
\end{figure}

This section demonstrates how the I-FENN (integrated finite element neural network) framework is implemented for the transient thermoelasticity problem. Although this paper focuses on using Seq2Seq learning models, where The Seq2Seq learning models developed here are based on the TCN architecture (see Figure \ref{tcn_fig}), we develop a PINN model for comparison purposes. The I-FENN framework is implemented in two steps. The first step is to train a PINN or PI-TCN model offline for the energy equation (see Equation \ref{Energy_eqn}) \cite{gurtin1973linear, abeyaratne1998continuum, truesdell2013linear} to predict the temperature profile based on the time, coordinates, and strain rate. The first step for the PINN and PI-TCN models is discussed in further detail below. The second step requires integrating the PI-TCN/PINN model inside the stiffness function at the element level, as illustrated in Figure \ref{I-FENN_fig}. Although this framework computationally decouples the energy equation from the linear momentum equation while performing the finite element analysis, its effect is still captured through the PI-TCN/PINN model. After training the PINN or PI-TCN model, it is integrated inside the finite element analysis, as depicted in Figure \ref{I-FENN_fig}. 

The PINN and PI-TCN models are trained based on the integration point data, as discussed later. Using higher-order elements for the same number of elements results in more integration points for training. This should potentially lead to more accurate results if the model is appropriately trained. However, this would require more computational resources and higher computational time. Post-training, the PINN and PI-TCN models offer a continuous function that maps inputs to outputs. This characteristic is particularly advantageous as it facilitates the model's application in predicting temperatures at both the integration points and the nodes, not just for the mesh used for training but also for finer meshes and/or higher-order discretizations. This aspect underscores the model's versatility and adaptability in various computational contexts. Investigating the scalability of the I-FENN framework in the context of extremely high-order elements or exploring optimization techniques to balance computational efficiency with accuracy could be valuable future endeavors.

In the current I-FENN implementation, the NN is used for capturing the thermal field, while the FEM is used for predicting the mechanical field. The suggestion to reverse this approach, applying the NN for the mechanical field and FEM for the thermal field, is indeed promising but lies outside the scope of this study, representing a potential avenue for future research. This alternative application presents certain complexities; mainly, using the NN for mechanical field necessitates handling vector outputs. The vectorial output of the mechanical field means that the NN must learn to predict multiple interrelated values simultaneously, which can be a more intricate task than scalar predictions. Furthermore, the components of displacement vectors might vary significantly in magnitude. This disparity poses a challenge for the NN training process, as it may lead to difficulties in converging to an accurate model. One might consider separate neural networks for different displacement components to address this. Alternatively, or additionally, dynamic scaling schemes could be employed. Such a scheme helps adaptively normalize the input and output data, which is especially useful when dealing with transient problems.

It is pertinent to underscore two key advantages of the I-FENN framework. Firstly, I-FENN offers a computational speed-up, which is crucial for the scalability of multiphysics problems, as this enables solving multiphysics problems with finer meshes and/or higher-order elements, as discussed above. Secondly, the framework paves the way for studying phenomena incorporating more physics, particularly three or more governing partial differential equations (PDEs). While the present study focuses on the mechanical and thermal fields, the framework should be applicable and may be more pronounced in more complex scenarios. For instance, if we consider a thermo-electric-mechanical problem, separate neural networks can be developed for distinct fields like thermal and electric. Subsequently, these two NN models are integrated within a finite element scheme for the mechanical field.

\subsection{Boundary conditions}\label{IFENN_BCs}

Dirichlet boundary conditions are enforced by subjecting the output layer of the neural network to:
\begin{equation}\label{distance_func}
\begin{aligned}
    T\left(t, x, y, \boldsymbol{\dot{\varepsilon}}, \boldsymbol{\phi}\right) &= A\left(x, y\right) + B\left(x, y\right) \circ \tilde{T}\left(t, x, y, \boldsymbol{\dot{\varepsilon}}, \boldsymbol{\phi}\right),\\   
\end{aligned}
\end{equation}
where $\boldsymbol {\phi}$ denotes the parameters of the neural network, $\boldsymbol {\phi}= \{\boldsymbol{W}, \boldsymbol{b}\}$. $\tilde{T}\left(t, x, y, \boldsymbol{\dot{\varepsilon}}, \boldsymbol{\phi}\right)$ is the output of the neural network before the enforcement of Dirichlet boundary conditions (DBCs), and $\circ$ denotes the Hadamard product between two vectors. $A\left(x, y\right)$ and $B\left(x, y\right)$ are distance functions chosen such that $T\left(t, x, y, \boldsymbol{\dot{\varepsilon}}, \boldsymbol{\phi}\right)$ satisfies the boundary conditions active on the physical domain, i.e., $A$ and $B$ are chosen such that:
\begin{equation}
\begin{aligned}
A\left(x, y\right) &= \overline{T}, \, \quad \forall (x, y) \in \Gamma_T,\\
B\left(x, y\right) &= 0, \, \, \quad \forall (x, y) \in \Gamma_T.
\end{aligned}
\end{equation}
The reader is referred to the work of Nguyen-Thanh et al. \cite{nguyen2020deep} and Rao et al. \cite{rao2021physics} for further details. The Neumann boundary conditions are imposed using the penalty method, as discussed in Section \ref{IFENN_Loss}. Specifically, we account for the Neumann boundary conditions by adding a term in the loss function definition imposing this type of boundary condition.

\subsection{Loss function}\label{IFENN_Loss}

Typically, to calculate the loss function and related variables, finding the solution's derivatives is required. Derivatives are computed through the use of automatic differentiation (AD), which is a feature available in deep learning frameworks. The unconstrained minimization problem reads as:
\begin{equation}\label{Minimization}
\begin{aligned}
\boldsymbol{\phi}^{*}&=\underset{\boldsymbol{\phi}}{\mathrm{arg\,min}} \; \mathcal{L}\left(\boldsymbol{\phi}\right){.}\\
\end{aligned}
\end{equation} 
where $\boldsymbol{\phi}^{*}$ denotes the optimized set of weights and biases.

The loss function $\mathcal{L}$ is defined using the residual norm ($L2$) in the light of the energy equation (see Equation \ref{new_energy}):
\begin{equation}\label{LOSS}
\begin{aligned}
    \mathcal{L}&=L2_E + \lambda_T \; L2_T + \lambda_q \; L2_q, \quad \text{where}\\
    L2_E &= \Vert \rho C_{\varepsilon} \dot{T} + \alpha (3\lambda+2\mu) T_{o} \text{tr}(\boldsymbol{\dot{\varepsilon}}) +  \boldsymbol{\nabla}\cdot \boldsymbol{q} \Vert^2{,}\\
    L2_T &= \Vert T_{FE} - T_{NN} \Vert^2{,}\\
    L2_q &= \Vert \boldsymbol{q}\cdot\boldsymbol{n} - \overline{q} \Vert^2,\\
\end{aligned}
\end{equation}
where $\Vert \cdot \Vert^2$ denotes the $L2$ norm. $\lambda_T >0$ and $\lambda_q>0$ are hyperparameters (penalty coefficients) weighting the contribution of the different loss terms. The numbers of the sampled points corresponding to the different terms of the loss function are $N_E$, $N_T$, and $N_q$.  Here, we set $N_E$, $N_T$, and $N_q$ to the total number of integration points in the finite element mesh. $L2_E$ accounts for the residuals of the energy equation inside the domain. $L2_T$ induces the network to minimize the discrepancy between the temperature predictions of the FE, $T_{FE}$, and neural network (NN), $T_{NN}$, models. $L2_q$ ensures that the NN model satisfies Neumann boundary conditions. 

It is crucial to specify the locations where each term of the loss function in Equation \ref{LOSS} is evaluated. In finite element analysis, the unknown temperatures are determined at the nodes, while the flux and strain rate are calculated at the integration points by utilizing the derivatives of the shape functions. During the training process, the $L2_E$ and $L2_T$ terms are evaluated at the integration points, where $T_{FE}$ is interpolated from the nodal temperature using the FE shape functions. It is worth mentioning that the nodal temperatures obtained from the finite element analysis are not used throughout the training process, i.e., they are not seen by the model during the training of the NN. However, they are used as a sanity check after the training to measure the quality of the NN results. The points used for the $L2_q$ term are equidistant along the boundary, $\Gamma_q$.

\subsection{I-FENN framework using PINN}\label{Section_ifenn_pinn}

\begin{figure}[!htb]
    \centering
    \includegraphics[width=0.9\textwidth]{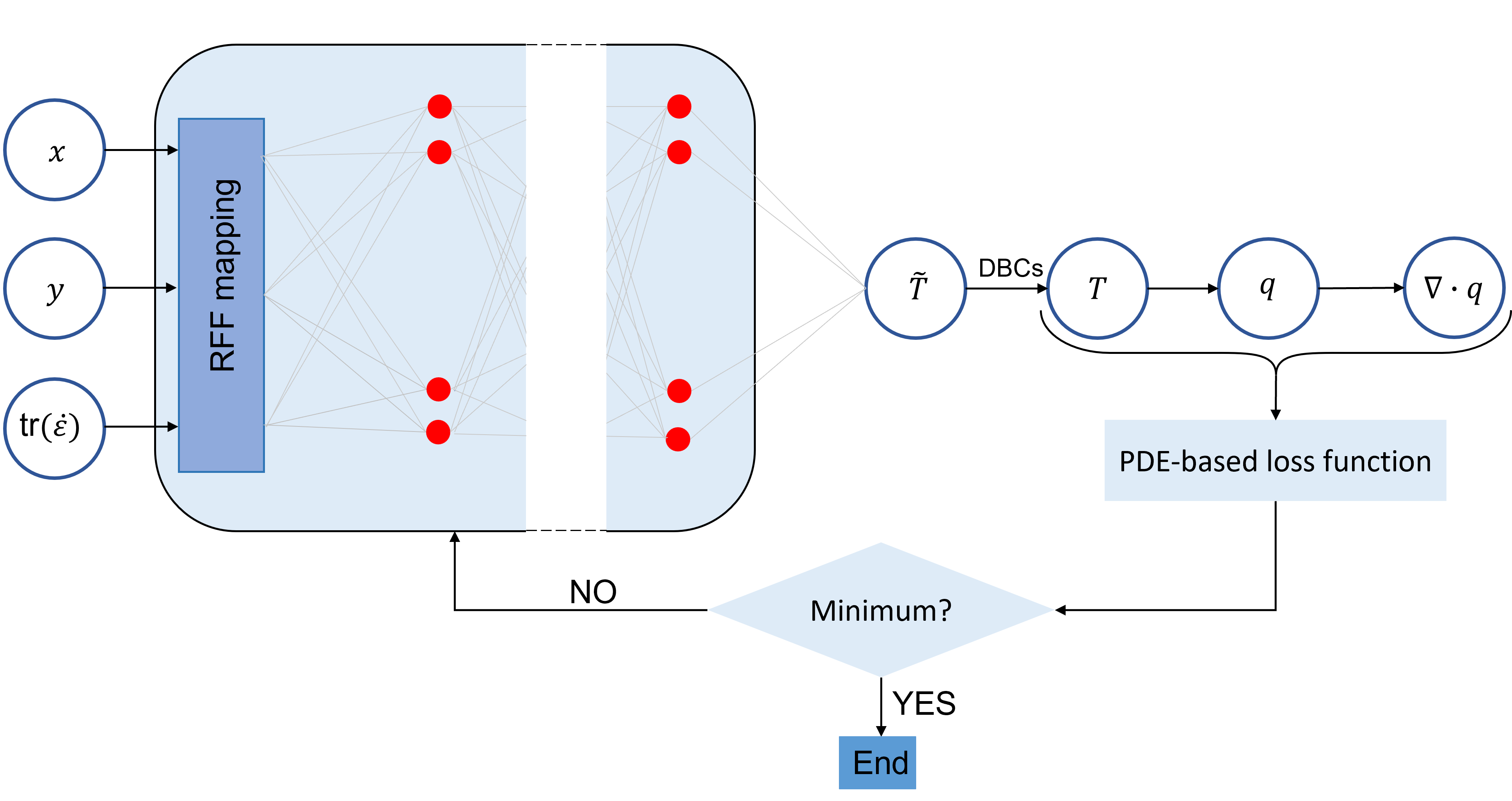}
    \caption{Training of the PINN model at $t_{n+1}$.}
    \label{mlp_I-FENN}
\end{figure}

Figure \ref{mlp_I-FENN} displays the training process of the PINN model. Different PINN models are trained at different time increments. The Euler discretization scheme is used for the MLP-based I-FENN framework, where the number of PINN models equals the number of time increments used to solve the problem, i.e., we have a PINN model for each time increment. The PINN models have three input features: $x$-coordinate, $y$-coordinate, and $\text{tr}(\boldsymbol{\dot{\varepsilon}})$. As shown in Figure \ref{mlp_I-FENN}, the three input variables are passed through a layer where Fourier feature mapping (see Section \ref{RFF}) is performed. Subsequently, the mapping outcome is passed through five consecutive neural layers. The number of neurons is a hyperparameter that determines the size of the network. The hyperbolic tangent (tanh) is used as an activation function for the PINN model.

Then, we impose the boundary conditions to the neural network output, $\tilde{T}$, as shown in Equation \ref{distance_func}, to obtain the temperature profile, $T$ at time increment $t_{n+1}$. The flux, $\boldsymbol{q}$, and the divergence of flux, $\boldsymbol{\nabla}\cdot\boldsymbol{q}$, are computed using automatic differentiation. Then, the loss function is obtained, as shown in Equation \ref{LOSS}. The process is repeated until convergence. For a fair comparison with the PI-TCN, one must consider PINN and PI-TCN models of similar size. More details are mentioned below in Section \ref{Plate_hole2D}.

\subsection{I-FENN framework using PI-TCN}\label{Section_ifenn_pitcn}

\begin{figure}[!htb]
    \centering
    \includegraphics[width=0.7\textwidth]{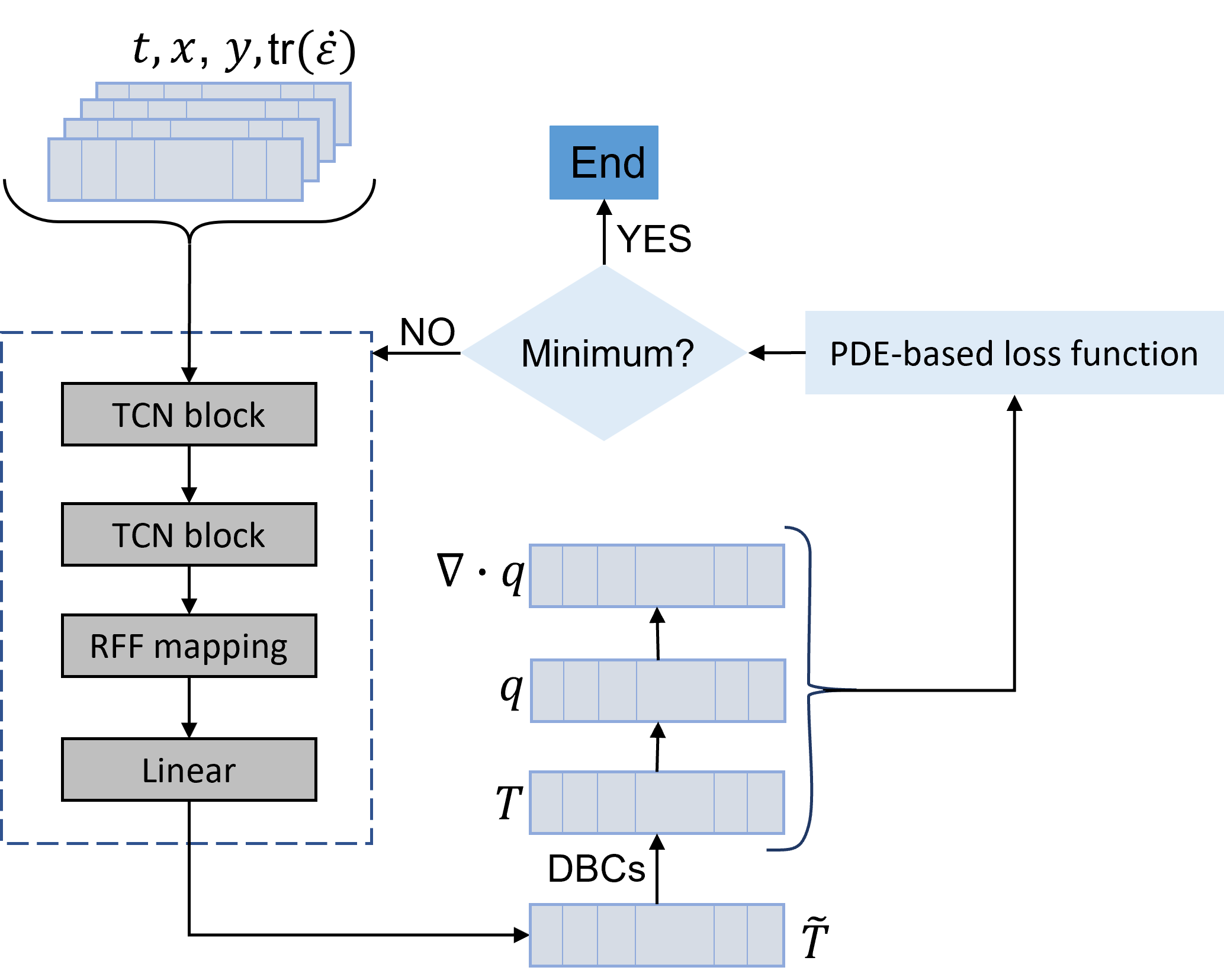}
    \caption{Training of the PI-TCN model.}
    \label{tcn_I-FENN}
\end{figure}

Figure \ref{tcn_I-FENN} depicts the training process of the PI-TCN model. Unlike the MLP-based I-FENN framework, for the TCN-based I-FENN framework, only one neural network is developed to predict temperatures at different time increments. TCN lies under the umbrella of Seq2Seq learning models, i.e., it supports sequential inputs and outputs. The inputs are cast into four sequences: time, $x$-coordinate, $y$-coordinate, and $\text{tr}(\boldsymbol{\dot{\varepsilon}})$, as shown in Figure \ref{tcn_I-FENN}. The time sequence is the same for all points in the domain. The $x$-coordinate and $y$-coordinate vary spatially to represent the position of a point in the domain, but they do not change temporally, i.e., they have the same value along their sequences for a certain point. The last input sequence is $\text{tr}(\boldsymbol{\dot{\varepsilon}})$, which varies spatially and temporally.

Figure \ref{tcn_I-FENN} shows that these sequential input variables are fed into two consecutive TCN blocks (see Figure \ref{tcn_fig} for the architecture of the TCN block). Then, the output is passed through a Fourier feature mapping and a fully connected (dense) layer. The output of the PI-TCN model is the $\tilde{T}$ sequence. Using Equation \ref{distance_func}, we ensure that the Dirichlet boundary conditions are fulfilled, and we obtain the $T$ sequence. With the help of automatic differentiation, we determine $\boldsymbol{q}$ and $\boldsymbol{\nabla} \cdot \boldsymbol{q}$; both are in sequential form.

\section{Numerical examples}\label{results}

This section demonstrates the proposed methodology via numerical implementation. Four different scenarios are examined and analyzed to assess the feasibility and potential of the I-FENN framework. First, in Section \ref{1D}, the I-FENN framework is applied to a 1D thermoelasticity model. The analysis is used to verify the I-FENN and PI-TCN model. The PI-TCN model is shown to outperform the data-driven TCN. In addition, the I-FENN's potential gains in computational time are presented. Then, Section \ref{plate2D} extends our study to a 2D domain that is clamped from the left-hand side and possesses prescribed temperatures on the top and bottom sides of the domain. In this example, we show that the I-FENN framework can solve 2D scenarios, and it also generalizes to refined meshes, although trained on a relatively coarse mesh. Additionally, we investigate how this approach leads to computational savings as more refined meshes are solved. Finally, Section \ref{Plate_hole2D} implements the I-FENN framework to a plate with a hole at the center, where a prescribed temperature is imposed on the hole boundary. Also, in Section \ref{Plate_hole2D}, we compare the performance of the conventional PINN and PI-TCN models and show that the PI-TCN model outperforms the PINN model. Section \ref{3DPlate} extends our study to 3D plate. This example demonstrates that the I-FENN framework can address 3D scenarios and extends to finer meshes despite being initially trained on a coarser mesh. Furthermore, the study examines the computational efficiency gained as increasingly refined meshes are processed. In all cases, we compare our results with those obtained solely from the finite element model of fully coupled thermoelasticity, which is discussed in Section \ref{FEA}. PI-TCN models are developed for the four cases mentioned above. We report the temperature variation, $\Theta=T - T_{o}$, rather than the absolute temperature, $T$, for the numerical examples discussed below. The material properties used for the numerical examples discussed below are $\lambda=40$GPa, $\mu=27$GPa, $\rho=2700$kg/m$^3$, $\alpha=2.31\times10^{-5}\text{m/(m °C)}$, $C_{\varepsilon}=910$J/(kg °C), and $k=237$W/(m °C). The reference temperature is $T_{o}=293$K. Two types of error measures are used in this paper to evaluate the performance of the I-FENN framework: absolute error, $Er_{abs}$, and relative error, $Er_{rel}$. The errors at a point $\boldsymbol{x}$ are expressed as:

\begin{equation}
\begin{aligned}
    Er_{abs}^{\boldsymbol{x}}(D) &=  |D_{I-FENN}^{\boldsymbol{x}} - D_{FEM}^{\boldsymbol{x}}|, \;\;\; \;\;\; \;\; \;\;\;\; \;\;\; \;\;\; \boldsymbol{x}\in\Omega \;\;\; \text{or} \;\;\; \boldsymbol{x}\in\Gamma,\\
    Er_{rel}^{\boldsymbol{x}}(D) &= \frac{|D_{I-FENN}^{\boldsymbol{x}} - D_{FEM}^{\boldsymbol{x}}|}{|D_{FEM}^{\boldsymbol{x}}|} \times 100\%,  \;\;\;\; \boldsymbol{x}\in\Omega \;\;\; \text{or} \;\;\; \boldsymbol{x}\in\Gamma,\\
\end{aligned}
\end{equation}
where the subscripts $I-FENN$ and $FEM$ represent the solution obtained using the I-FENN framework and the fully coupled finite element analysis, respectively. $D$ is the field for which the error is computed, i.e., $D$ is the temperature variation, $\Theta$, the displacement along the $x-$axis, $u$, the displacement along the $y-$axis, $v$, or the displacement along the $z-$axis, $w$. The hyperparameters of the PI-TCN model used in the four examples are summarized in Table \ref{PITCN_hyperparams}.

\begin{table}[]
\centering
\caption{Hyperparameters of the PI-TCN model.}
\label{PITCN_hyperparams}
\begin{tabular}{|c|c|}
\hline
Number of filters       & 16           \\ \hline
Number of hidden layers & 4            \\ \hline
Dilation rates, $d_i$   & [1, 2, 4, 8] \\ \hline
Kernel size, $k_{size}$ & 11           \\ \hline
RFF $\sigma$            & 0.07         \\ \hline
$\lambda_T$             & 1.0          \\ \hline
$\lambda_q$             & 200          \\ \hline
\end{tabular}
\end{table}

The hyperbolic tangent (tanh) function is adopted as the activation function for both the PINN and PI-TCN models developed in this paper. The proposed framework is executed in PyTorch \cite{NEURIPS2019_9015}, where the network parameters $\boldsymbol{\phi}$ are optimized using the limited Broyden–Fletcher–Goldfarb–Shanno (LBFGS) algorithm with Strong Wolfe line search \cite{liu1989limited, lewis2013nonsmooth}. Once the NN is trained, as discussed above, the trained model is used to predict the nodal temperatures, which are needed for evaluating the mechanical weak form (see Equation \ref{mech_weak}).

\subsection{1D thermoelasticity}\label{1D}

\begin{figure}[!htb]
    \centering
    \includegraphics[width=0.9\textwidth]{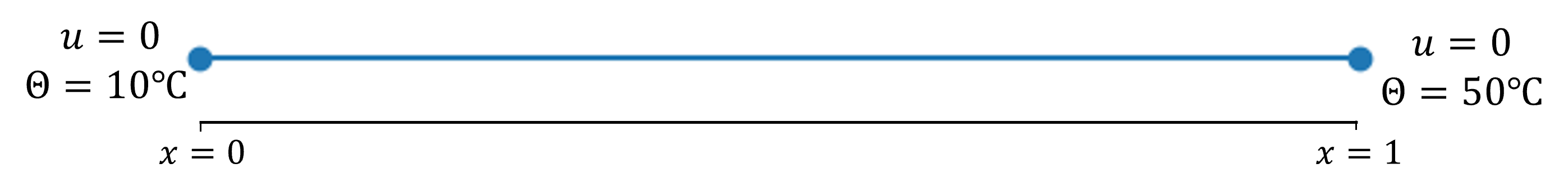}
    \caption{Schematic illustration of the geometry and boundary conditions of the 1D thermoelasticity problem.}
    \label{1D_example}
\end{figure}

The first numerical example is a 1D domain with a unity length, as shown in Figure \ref{1D_example}. A zero displacement and a temperature variation of $\Theta=10\text{°C}$ are imposed on the left boundary, while a zero displacement and a temperature variation of $\Theta=50\text{°C}$ are applied on the right boundary. 

\subsubsection{I-FENN results}\label{prelim}
\begin{figure}[!htb]
    \centering
    \includegraphics[width=1.0\textwidth]{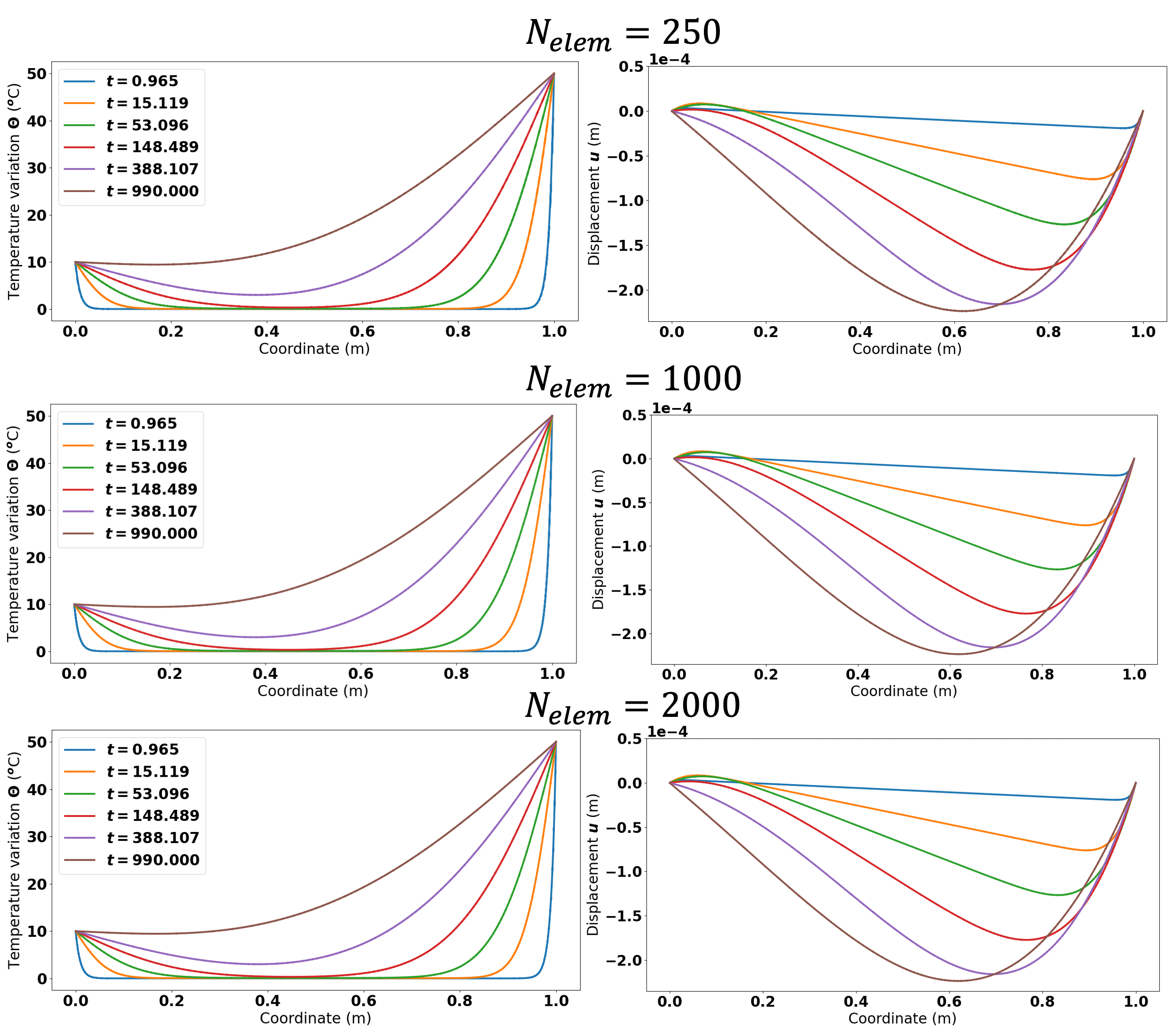}
    \caption{Finite element solution for three different mesh sizes.}
    \label{fe_mesh}
\end{figure}

A fully coupled finite element analysis is first performed, as described in Section \ref{FEA}, to generate the training data. The finite element analysis uses 50-time increments on the logarithmic scale. Figure \ref{fe_mesh} displays the temperature variation and displacement for three different mesh sizes at different time increments, where $N_{elem}$ denotes the number of elements. Similar and consistent time-dependent responses are observed for all mesh sizes.

\begin{figure}[!htb]
    \centering
    \includegraphics[width=1.0\textwidth]{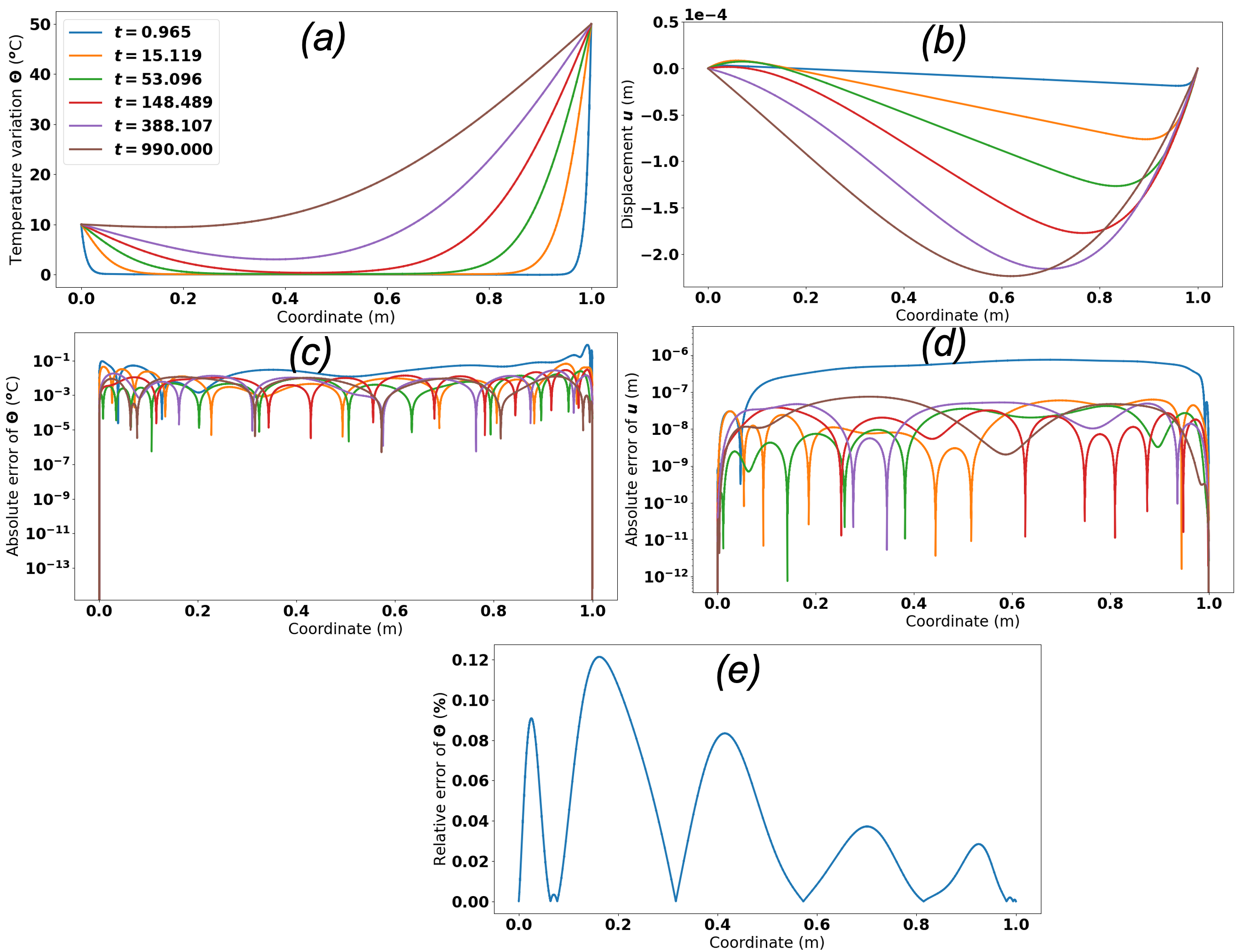}
    \caption{Solution obtained using the I-FENN framework: a) temperature variation $\Theta$ obtained from the PI-TCN model, b) displacement, c) $Er_{abs}^{\boldsymbol{x}}(\Theta)$, d) $Er_{abs}^{\boldsymbol{x}}(u)$, and e) $Er_{rel}^{\boldsymbol{x}}(\Theta)$ at $t=990$s. The figure legend shown in (a) is applicable to (b), (c), and (d).}
    \label{1d_I-FENN_results}
\end{figure}

Next, the problem is solved using the proposed I-FENN framework. A mesh size of $N_{elem}=2000$ is utilized. As discussed in Section \ref{I-FENN}, we develop a PI-TCN model for the energy equation (see Equation \ref{new_energy}). In this example, there are no flux boundary conditions; therefore, Equation \ref{LOSS} reduces to the first two terms: $L2_E$ and $L2_T$. Then, the PI-TCN model is integrated inside a finite element model to determine the temperature required to calculate the mechanical weak form. Figure \ref{1d_I-FENN_results} summarizes the results attained using the I-FENN framework. Figure \ref{1d_I-FENN_results}a shows the temperature variation obtained from the trained PI-TCN model at the nodes of the finite element mesh. As mentioned in Section \ref{I-FENN}, the PI-TCN model is trained using integration point values, not nodal values of variables. Hence, the error calculations shown in Figure \ref{1d_I-FENN_results} account for the errors induced by the interpolation between the integration points and nodes. Figures \ref{1d_I-FENN_results}c and \ref{1d_I-FENN_results}d depict the absolute errors of the temperature variation $\Theta$ and displacement $u$, respectively. The fully coupled finite element model results are used as a reference for error quantification. It is observed that the highest $Er_{abs}^{\boldsymbol{x}}(\Theta)$ and $Er_{abs}^{\boldsymbol{x}}(u)$ take place at the first time increment, $t=0.965$s. Figure \ref{1d_I-FENN_results}e shows $Er_{rel}^{\boldsymbol{x}}(\Theta)$ at the last time increment, $t=990{s}$. The highest $Er_{rel}^{\boldsymbol{x}}(\Theta)$ is around $0.12\%$. Furthermore, the findings shown in Figure \ref{1d_I-FENN_results} indicate that Equation \ref{distance_func} has enabled strict fulfillment of the Dirichlet boundary conditions. The use of Equation \ref{distance_func} defines the "admissible" function space of the solution.

Figure \ref{1d_I-FENN_results} shows that the errors exhibit oscillating behavior. The solution obtained using the I-FENN framework is compared against the fully coupled finite element method, which is a numerical solution. Several potential factors might be at play. The interpolation process between integration points and nodes, coupled with the use of automatic differentiation in the computation of spatial gradients, is likely to introduce variances in the solutions, which in turn could skew the characterization of error. Moreover, convergence thresholds set within iterative solvers may yield minor but noticeable oscillations in error as the solution nears convergence. Additionally, the implementation of boundary conditions, which differs between the two methods, may be a source of distinct solutions at the periphery, potentially leading to the observed oscillatory error patterns. Furthermore, the PI-TCN model, despite being trained on the integration points of an FE coarse mesh, establishes a continuous mapping between inputs and outputs. This factor could also influence the results. These points are discussed further in Section \ref{conclu}.

\begin{figure}[!htb]
    \centering
\includegraphics[width=0.4\textwidth]{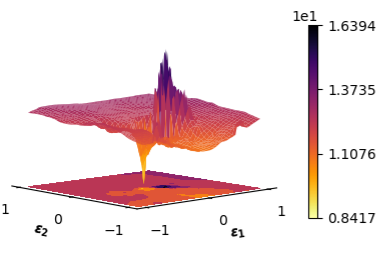}
    \caption{Loss landscapes of the PI-TCN model for the 1D example (Section \ref{prelim}).}
    \label{1d_losslandscape}
\end{figure}

At this point, we visually represent the loss landscapes for our trained PI-TCN model to gain insights into the difficulties associated with training PI-TCN models. 
Figure \ref{1d_losslandscape} shows the two-dimensional loss landscapes of the PI-TCN model. The loss landscapes display a rugged and uneven terrain, with an apparent minimum point to which the PI-TCN model converges. Similar observations are noted in the work of Basir \cite{basir2023investigating} when the LBFGS optimizer is used to solve other physical phenomena using PINN (MLP-based) models. This is because the LBFGS optimizer integrates the curvature data of the objective function throughout the training procedure \cite{basir2023investigating}. Figure \ref{1d_losslandscape} implies that it is challenging to navigate such landscapes while training the PI-TCN model. More research is still to be done on understanding the nature of the optimization process of PI-TCN models and other physics-informed machine learning models.

\subsubsection{Steady-state solution}\label{steady_state}
\begin{figure}[!htb]
    \centering
    \includegraphics[width=1.0\textwidth]{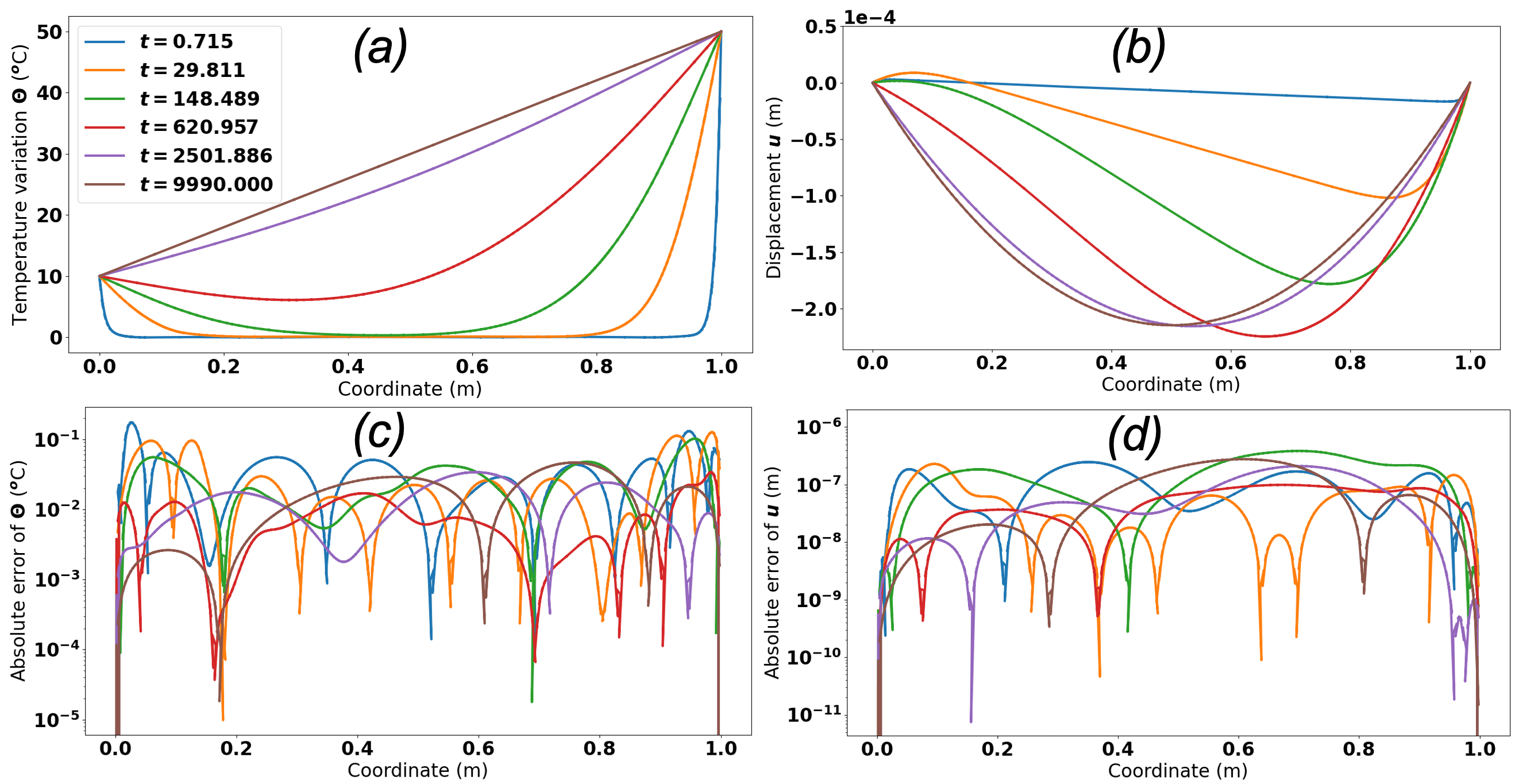}
    \caption{Solution obtained using the I-FENN framework when 250 elements and 100-time increments are considered. a) temperature variation $\Theta$ obtained from the PI-TCN model, b) displacement, c) $Er_{abs}^{\boldsymbol{x}}(\Theta)$, and d) $Er_{abs}^{\boldsymbol{x}}(u)$. The figure legend shown in (a) is applicable to (b), (c), and (d).}
    \label{1d_250elem_100Tincr}
\end{figure}

To further show the capabilities of the I-FENN framework, we consider a coarser mesh, $N_{elem}=250$, and attempt to capture the near steady-state solution. Hence, the problem is solved until a total time of around $10000$s in 100-time increments using logarithmic time incrementation. Like the previous showcases, the PI-TCN is initially trained to predict the temperature variation, $\Theta$, across the domain for all the time increments. Then, the trained PI-TCN is integrated inside the finite element scheme to obtain the displacement field. Figure \ref{1d_250elem_100Tincr} shows the solution obtained using the I-FENN framework and displays the errors computed against the solution obtained from the fully coupled finite element method. The comparison implies that both solutions are in agreement. Also, Figure \ref{1d_250elem_100Tincr} indicates that the PI-TCN and I-FENN framework scales well in a larger total time, and it captures the near steady-state solution, even with a small number of elements. 

\subsubsection{I-FENN: PI-TCN versus data-driven TCN}\label{DDTCNvsPITCN}
\begin{figure}[!htb]
    \centering
    \includegraphics[width=1.0\textwidth]{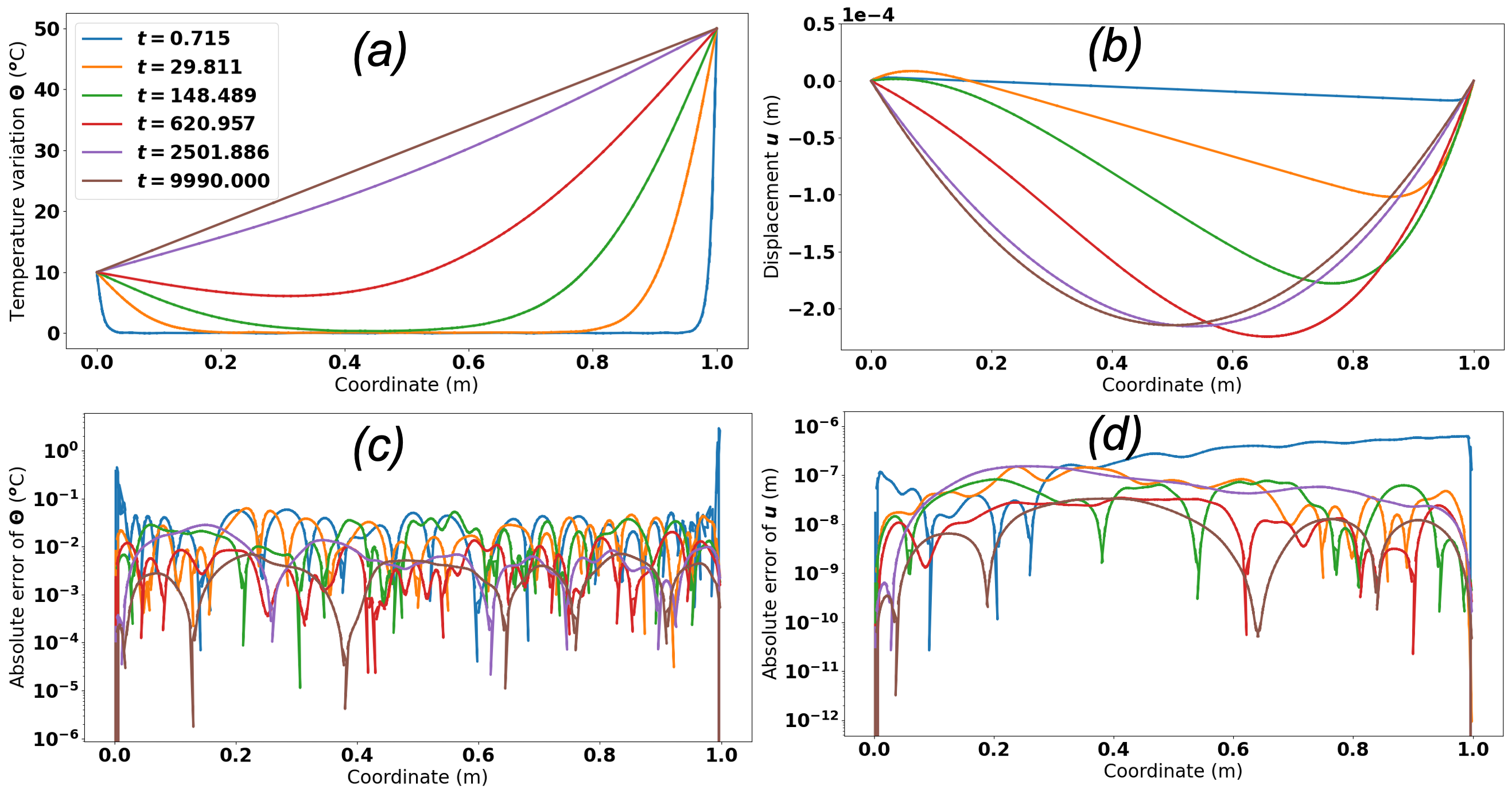}
    \caption{Solution obtained using the I-FENN framework using a data-driven approach without incorporating physical laws. 250 elements and 100-time increments are considered. a) temperature variation $\Theta$ obtained from the data-driven model, b) displacement, c) $Er_{abs}^{\boldsymbol{x}}(\Theta)$, and d) $Er_{abs}^{\boldsymbol{x}}(u)$. The figure legend shown in (a) is applicable to (b), (c), and (d).}
    \label{1d_250_data}
\end{figure}

\begin{figure}[!htb]
    \centering
\includegraphics[width=0.8\textwidth]{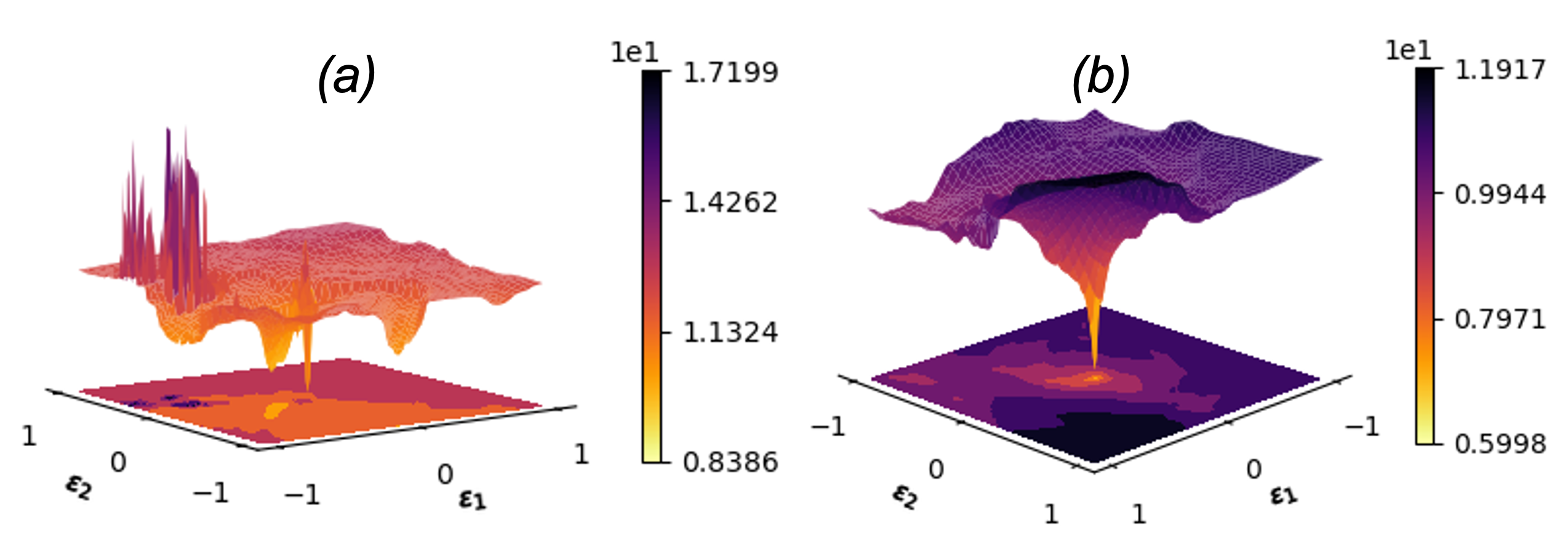}
    \caption{Comparison between the loss landscapes of the a) PI-TCN model and b) data-driven TCN model (see Section \ref{DDTCNvsPITCN}).}
    \label{1d_pitcn_vs_DDTCN_landscapes}
\end{figure}

Next, we show how the PI-TCN model trained based on the energy equation and data generated from the finite element (see Equation \ref{LOSS}) compares with the purely data-driven approach. Since this example does not have Neumann boundary conditions, the loss function for the PI-TCN has only two terms: $L2_E$ and $L2_T$, while the data-driven TCN model has only one term, which is $L2_T$. Please note that these two models have the same architecture shown in Figure \ref{tcn_I-FENN}, and the only difference is the definition of the loss function. Let us consider the same settings discussed in Section \ref{steady_state}, i.e., the number of elements is $N_{elem} =250$, and the total time is around $10000$s, solved in 100-time increments using logarithmic time incrementation. The data-driven TCN model is initially trained, and then it is integrated inside the finite element scheme to find the displacement field. Figure \ref{1d_250_data} depicts the results obtained using the data-driven machine learning model. In general, the I-FENN framework based on the data-driven TCN model can capture the temperature and displacement. However, higher errors are observed compared to the solutions attained when the PI-TCN is considered, as shown in the error plots of Figures \ref{1d_250elem_100Tincr} and \ref{1d_250_data}. Although the PI-TCN model induces lower errors than the data-driven TCN model, the training of the PI-TCN is more challenging, as it is implied from the loss landscapes shown in Figure \ref{1d_pitcn_vs_DDTCN_landscapes}. The loss landscapes of the PI-TCN are more rugged and less smooth than the data-driven TCN model. Also, the loss landscapes of the PI-TCN show multiple distinct local minima. 

\subsubsection{Training on a coarse mesh and predicting on a fine mesh}\label{1d_coarse_to_fine}
 
\begin{figure}[!htb]
    \centering
\includegraphics[width=1.0\textwidth]{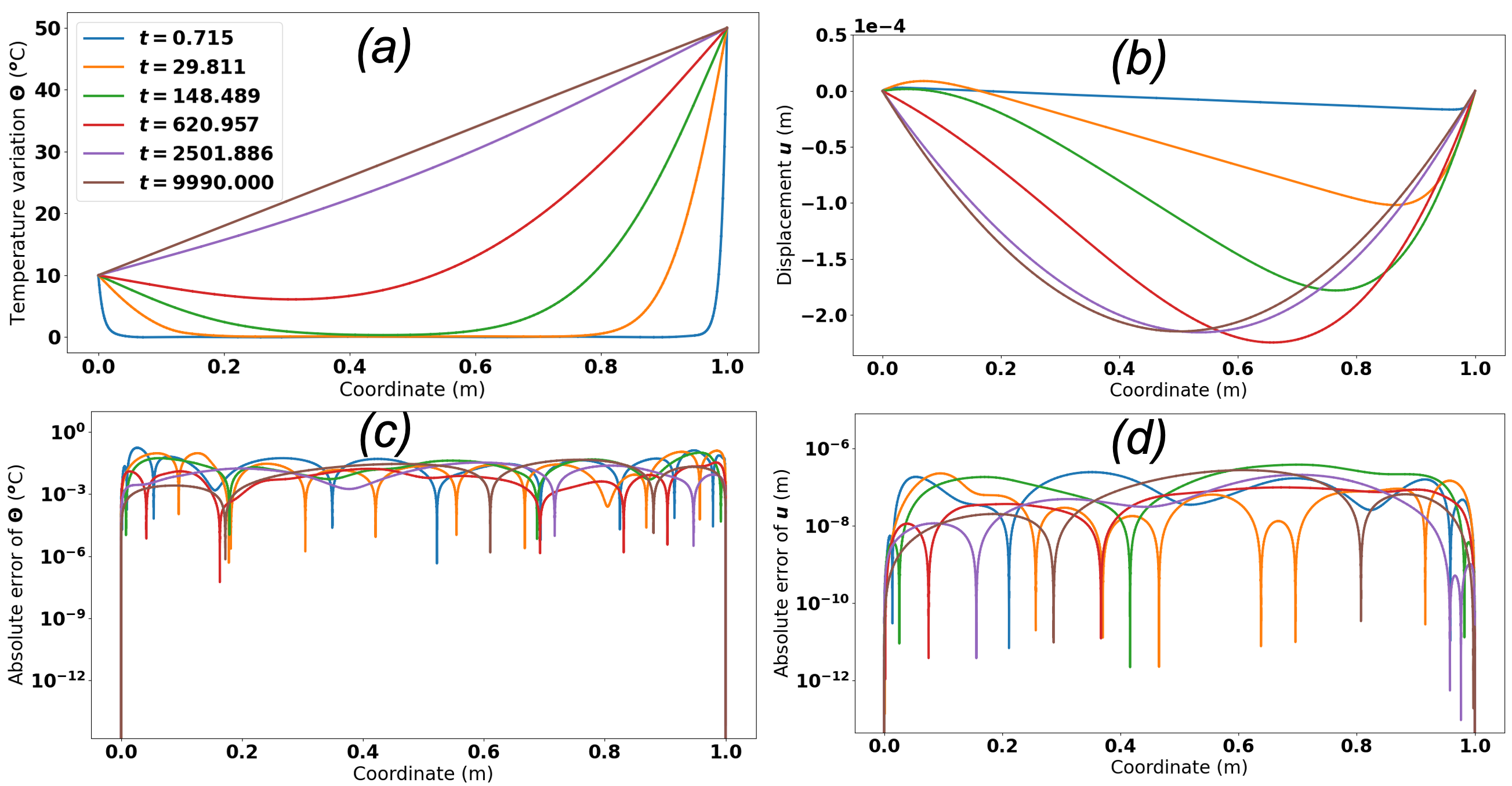}
    \caption{Comparison between the results obtained using the I-FENN framework and finite element method, when the PI-TCN model is trained using the coarse mesh, $N_{elem}=250$, and provides the nodal $\Theta$ for the fine mesh, $N_{elem}=10000$. Then, the trained PI-TCN model is integrated inside the mechanical weak form to obtain the displacements for the fine mesh. a) $\Theta$ from the I-FENN framework, b) $u$ from the I-FENN framework, c) $Er_{abs}^{\boldsymbol{x}}(\Theta)$, d) $Er_{abs}^{\boldsymbol{x}}(u)$.} 
    \label{1d_fine_mesh}
\end{figure}

Now, we examine the ability of the trained PI-TCN model to predict the solution of finer meshes. Adopting the same settings used in Section \ref{steady_state}, the PI-TCN is trained on a relatively coarser mesh having $N_{elem}=250$ elements, using 100-time increments on the logarithmic scale. Then, it is integrated inside the finite element scheme to obtain $\Theta$ for a refined mesh with $N_{elem}=10000$ elements (see Figure \ref{1d_fine_mesh}a). Subsequently, $\Theta$ computed from the PI-TCN is used to solve the mechanical weak form and obtain the displacement field for the refined mesh (see Figure \ref{1d_fine_mesh}b).

Figures \ref{1d_fine_mesh}c and \ref{1d_fine_mesh}d illustrate the absolute error of the I-FENN framework, where the PI-TCN is trained on a coarse mesh and used for obtaining the solution for the refined mesh. The reference solution for error calculations is obtained from the fully coupled finite element method with the refined mesh. When we compare the errors observed here with those reported for the case when the PI-TCN is trained on the coarse mesh and used to predict $\Theta$ also for the coarse mesh, as shown in Figure \ref{1d_250elem_100Tincr}, we observe similar error values. This implies that the PI-TCN maintains the same level of accuracy across meshes and does not cause any error magnifications due to transitioning to finer meshes. Hence, we infer that the PI-TCN model can generalize to meshes other than the one used for training.

\subsubsection{Computational time}\label{1d_comput_time} 

We compared the computational time of the I-FENN framework and the fully coupled finite element analysis. To obtain $\Theta$ for different mesh sizes, we trained the PI-TCN on a coarse mesh of 100 elements and used 300-time increments, which took 935 seconds. We then integrated it into the finite element scheme and utilized $\Theta$ to solve the mechanical weak form, resulting in the displacement field for the refined meshes. 

\begin{figure}[!htb]
    \centering
\includegraphics[width=0.8\textwidth]{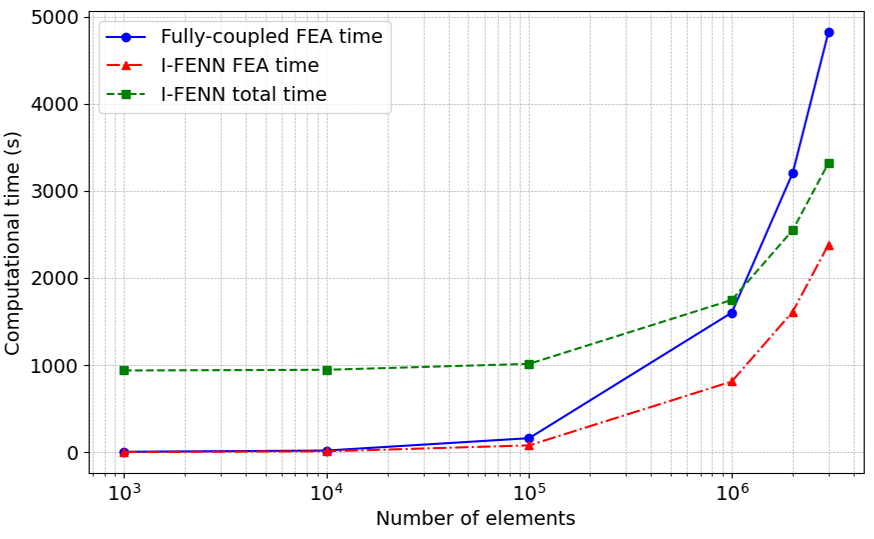}
    \caption{Computational time comparison between the fully coupled finite element analysis and the I-FENN framework.} 
    \label{1d_comp_time}
\end{figure}

Figure \ref{1d_comp_time} illustrates the computational time necessary for the finite element analysis using varying mesh sizes. Additionally, the figure presents two computational times for the I-FENN framework, including and excluding the PI-TCN training time. Although the finite element analysis requires less time at a small number of elements, the I-FENN framework exhibits superior efficiency as models scale to a significantly large number of elements, even when accounting for the training time. 

It is essential to highlight that our methodology's primary goal is to train a network once and use it multiple times. This approach has been extensively used in many other studies \cite{jokar2021finite, liang2018deep, mozaffar2019deep}. This aspect is particularly significant in industrial applications where identical simulations are frequently utilized for an extended period. Our approach provides a streamlined and efficient solution, offering substantial benefits in such scenarios. For example, studies conducted by Lu et al. \cite{lu2017analysis}, Moon et al. \cite{moon2015effect}, and Kadir et al. \cite{kadir2019computational} have examined thermal stress and temperature fields in different mechanical systems. Implementing our approach could significantly improve computational efficiency for the analyses performed in these studies.

\begin{figure}[!htb]
    \centering
\includegraphics[width=0.85\textwidth]{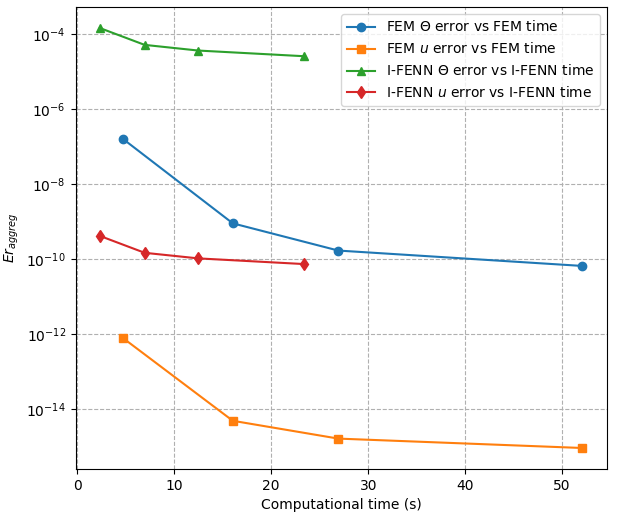}
    \caption{Solution error vs. computational time for the FEM and I-FENN framework.}
    \label{error_vs_comp_time_1d_convergence}
\end{figure}

Before concluding this 1D example, we use different meshes to investigate how the error evolves with the computational time for the solutions obtained from the fully coupled finite element and I-FENN framework. We use the fully coupled finite element analysis using the monolithic approach as our reference solution, with the acknowledgment that it is a numerical solution and is not the exact solution. However, the reference solution has a relatively finer solution. Therefore, it can be deemed as a more accurate solution based on the well-established knowledge of the convergence of the FEM solution. In other words, we plot the computational time for the I-FENN framework and FE against the solution error, using the finest FE solution as the reference solution (see Figure \ref{error_vs_comp_time_1d_convergence}). To show how the error converges with finer meshes, corresponding to higher computational time, we use the following error metric:
\begin{equation}
\begin{aligned}
    Er^{i}(D) &= \frac{1}{N_{nodes}} \left\lVert {D}^{i}_{IFENN}(\boldsymbol{x}) - {D}^{i}_{FEM}(\boldsymbol{x}) \right\rVert^{2}, \quad \boldsymbol{x} \in \Omega \text{ or } \boldsymbol{x} \in \Gamma,\\
    Er^{D}_{aggreg} &= \frac{1}{N_{inc}} \left\lVert Er^{i}(D) \right\rVert^{2},\\
\end{aligned}
\end{equation}
where the superscript $i$ indicates the time increment at which the error is computed. The error $Er^{i}$ measures the error across the entire field at a particular time increment, and it is normalized by the number of nodes $N_{nodes}$ in the domain. $D$ serves as the field for which the error is determined. Taking the norm of $Er(D)$ and normalizing it by the number of time increments $N_{inc}$ yield the aggregated error $Er^{D}_{aggreg}$. The reference solution is obtained from a monolithic finite element method with $N_{elem}=64,000$ elements. Then, the four points appearing in each curve correspond to the following meshes: $N_{elem}=1,000$, $N_{elem}=8,000$, $N_{elem}=16,000$, and $N_{elem}=32,000$ elements. Figure \ref{error_vs_comp_time_1d_convergence} shows that the I-FENN framework is less accurate than the FEM. The trends shown in this figure are expected. The neural network empowering I-FENN is trained using data from the coarsest FEM mesh; therefore, the consequent I-FENN accuracy will be capped at the accuracy of the coarsest FEM solution.
 
\subsection{2D plate}\label{plate2D}

\begin{figure}[!htb]
    \centering
    \includegraphics[width=0.5\textwidth]{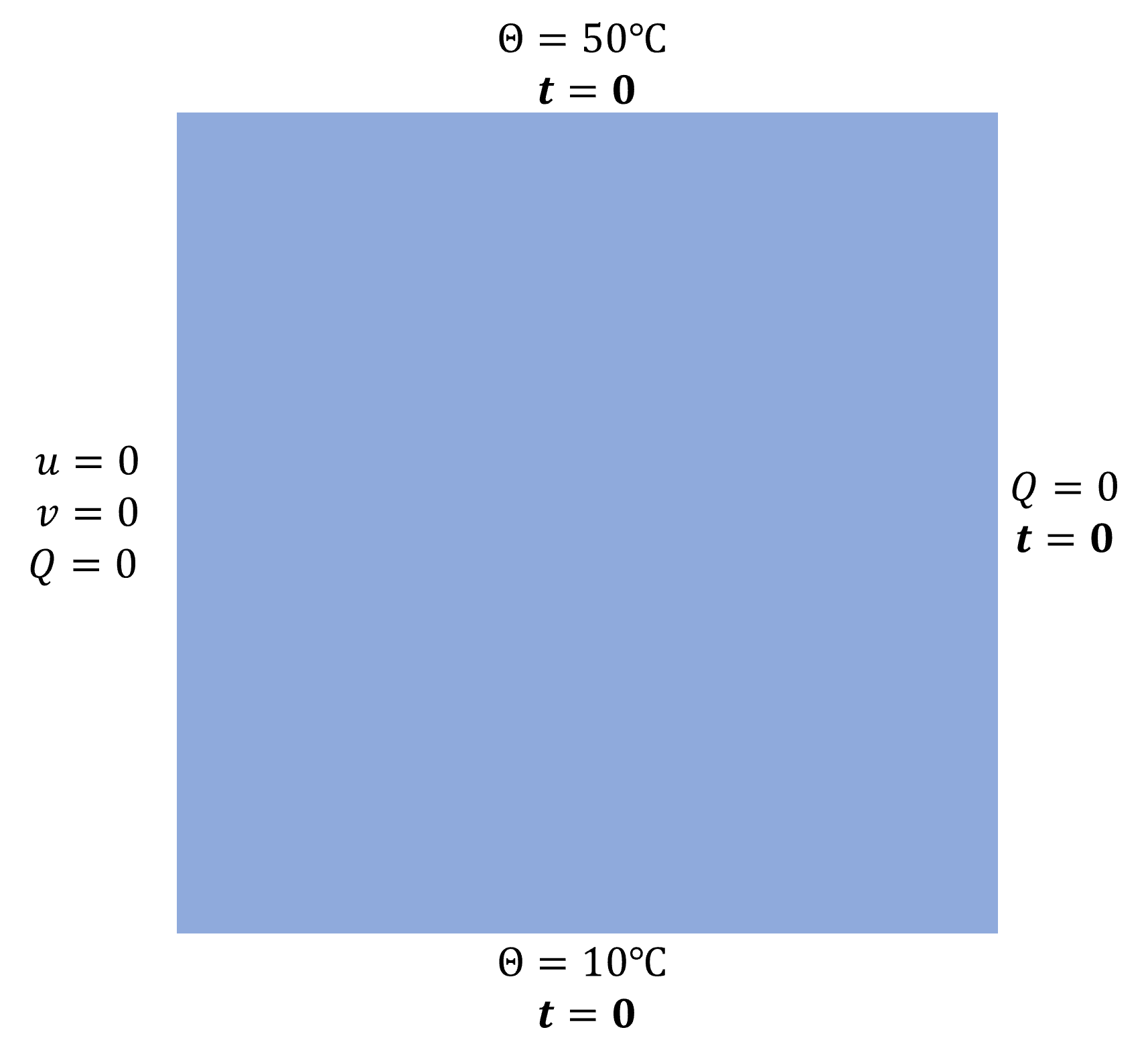}
    \caption{Schematic illustration of the geometry and boundary conditions of the 2D plate problem.}
    \label{2D_example}
\end{figure}

\begin{figure}[!htb]
    \centering
    \includegraphics[width=1.0\textwidth]{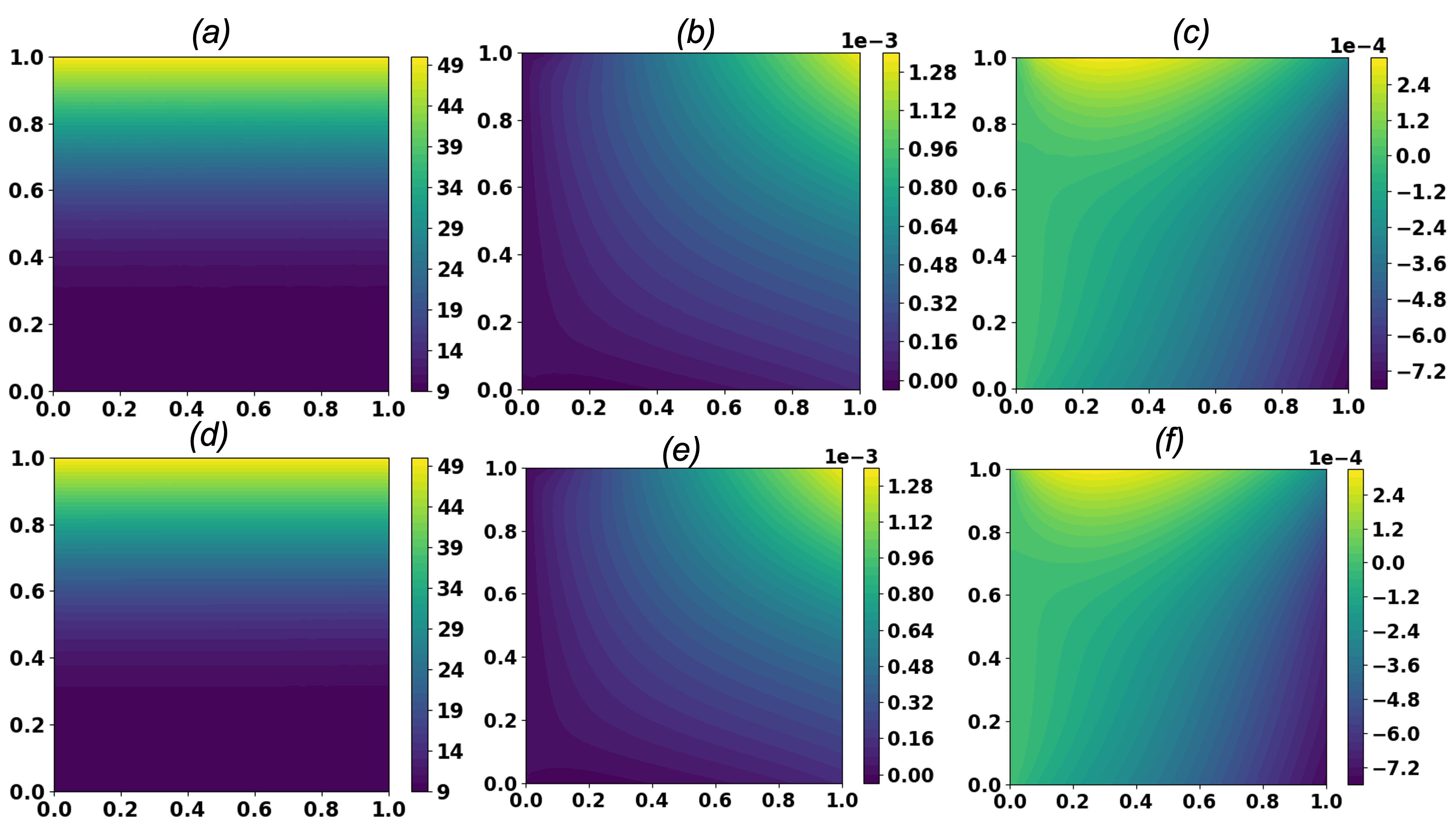}
    \caption{Finite element solutions for two different mesh sizes: a) $\Theta$ for $N_{elem}=702$, b) $u$ for $N_{elem}=702$, c) $v$ for $N_{elem}=702$, d) $\Theta$ for $N_{elem}=7722$, e) $u$ for $N_{elem}=7722$, and f) $v$ for $N_{elem}=7722$.}
    \label{2D_plate_mesh}
\end{figure}

The following numerical example is a square plate with sides of unit length. Figure \ref{2D_example} depicts the geometry and boundary conditions used in this example. A fully coupled finite element model is first developed to generate the training data. Fifty logarithmically spaced time increments are utilized in the finite element analysis. We start with a mesh dependence study; the problem is solved using a different number of elements: $N_{elem}=702$, $N_{elem}=1185$, $N_{elem}=2812$, and $N_{elem}=7722$. We observe that the results are mesh-independent. For brevity, we just include a comparison between the results when $N_{elem}=702$ and $N_{elem}=7722$ are used, as shown in Figure \ref{2D_plate_mesh}. 

\begin{figure}[!htb]
    \centering
\includegraphics[width=0.4\textwidth]{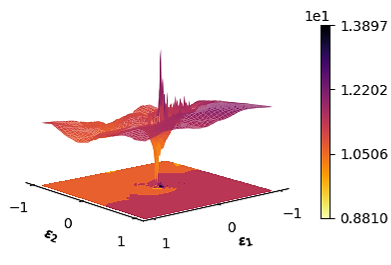}
    \caption{Loss landscapes of the PI-TCN model for the 2D plate example.}
    \label{2d_losslandscape}
\end{figure}

\begin{figure}[!htb]
    \centering
\includegraphics[width=1.0\textwidth]{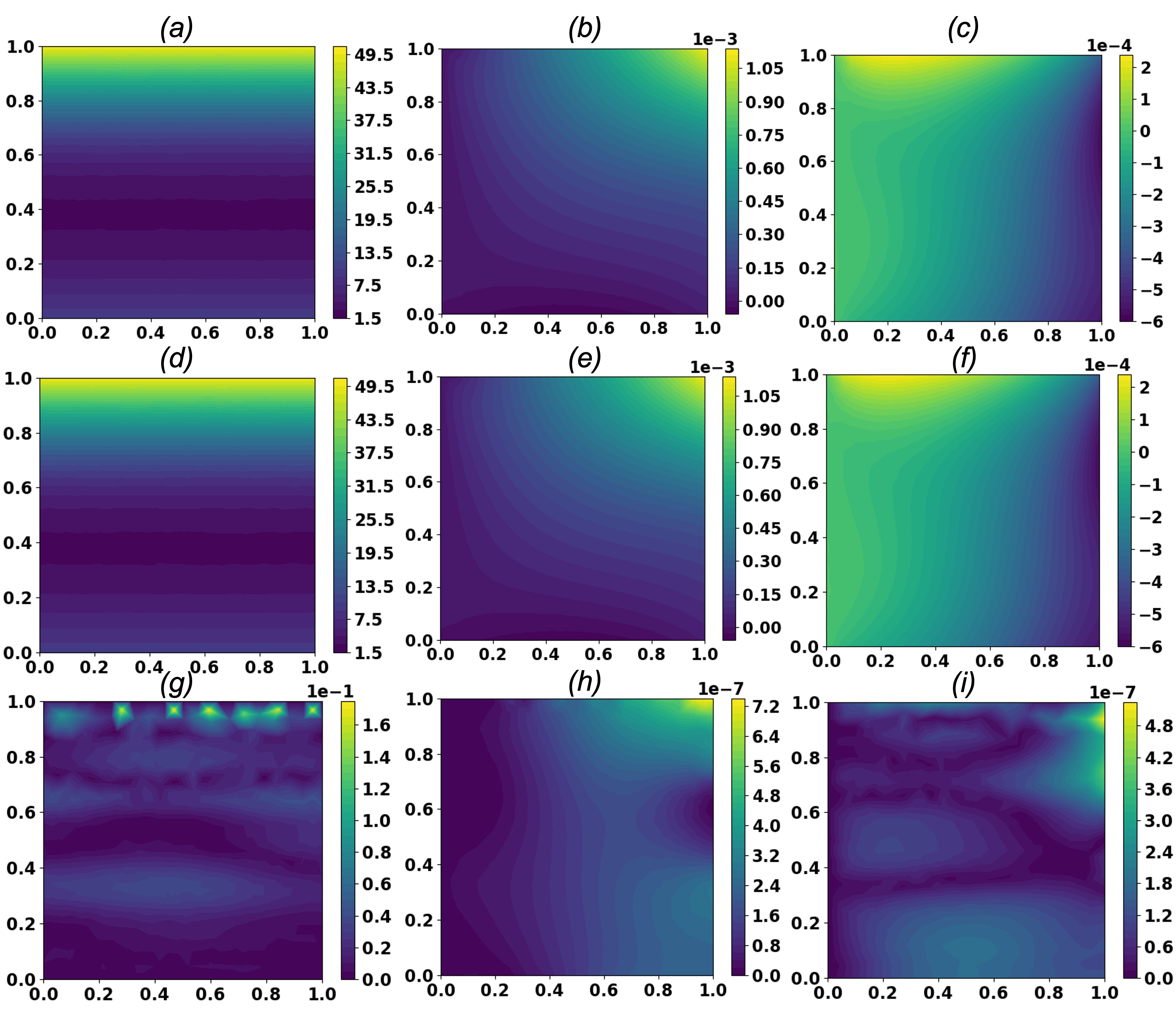}
    \caption{Comparison between the results obtained using the I-FENN framework and finite element method at $t=388.1$s. Temperatures are in °C, and displacements are in m. a) $\Theta$ from the I-FENN framework, b) $u$ from the I-FENN framework, c) $v$ from the I-FENN framework, d) $\Theta$ using the FEM, e) $u$ using the FEM, f) $v$ using the FEM, g) $Er_{abs}^{\boldsymbol{x}}(\Theta)$, h) $Er_{abs}^{\boldsymbol{x}}(u)$, and i) $Er_{abs}^{\boldsymbol{x}}(v)$.} 
    \label{2d_t39}
\end{figure}

\begin{figure}[!htb]
    \centering
\includegraphics[width=1.0\textwidth]{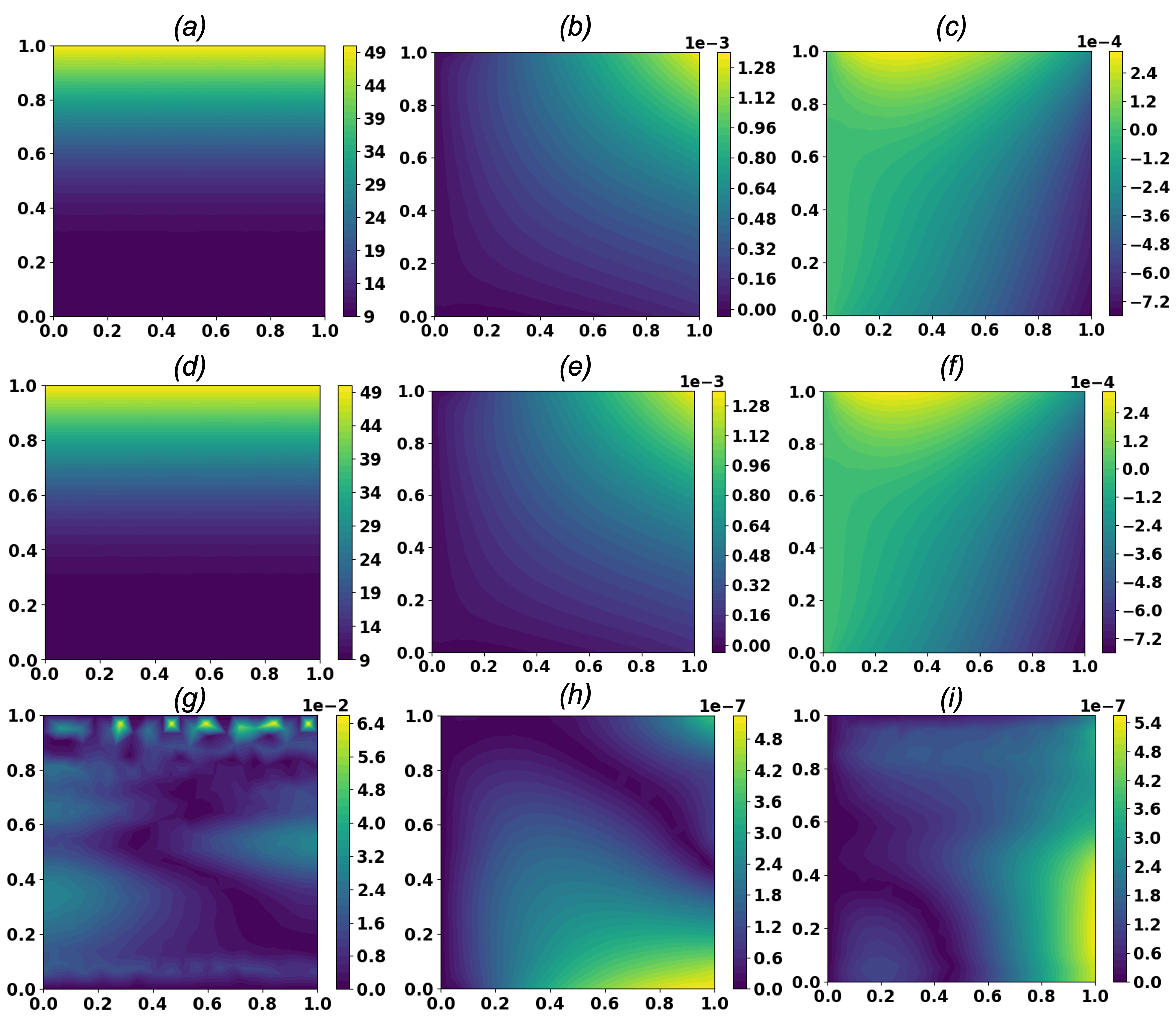}
    \caption{Comparison between the results obtained using the I-FENN framework and finite element method at $t=990$s. Temperatures are in °C, and displacements are in m. a) $\Theta$ from the I-FENN framework, b) $u$ from the I-FENN framework, c) $v$ from the I-FENN framework, d) $\Theta$ using the FEM, e) $u$ using the FEM, f) $v$ using the FEM, g) $Er_{abs}^{\boldsymbol{x}}(\Theta)$, h) $Er_{abs}^{\boldsymbol{x}}(u)$, and i) $Er_{abs}^{\boldsymbol{x}}(v)$.} 
    \label{2d_t49}
\end{figure}

Next, we train our PI-TCN model that captures the energy equation. A mesh size of $N_{elem}=702$ is used to train the PI-TCN model. Similar to the previous example, we visualize the loss landscapes. Figure \ref{2d_losslandscape} shows that the loss landscapes possess non-smooth terrain reflecting the challenges in training such a PI-TCN model. Subsequently, the finite element model incorporates the PI-TCN model to ascertain the temperature necessary for the weak form of the balance of linear momentum. Figures \ref{2d_t39} and \ref{2d_t49} compare the I-FENN framework and FEM results at different time increments. The results obtained using the fully coupled finite element model are used as a reference for error computations. It is observed that the results obtained from the I-FENN are in agreement with those obtained using the FEM. However, the temperature profile exhibits relatively higher errors that oscillate near the top boundary (see Figures \ref{2d_t39}g and \ref{2d_t49}g). Also, we highlight that the highest absolute errors are observed near boundaries. This numerical instability is a persistent challenge that many researchers have discussed \cite{appadu2016computational, liu2010stabilized, gu2017effective}. 

\begin{figure}[!htb]
    \centering
\includegraphics[width=1.0\textwidth]{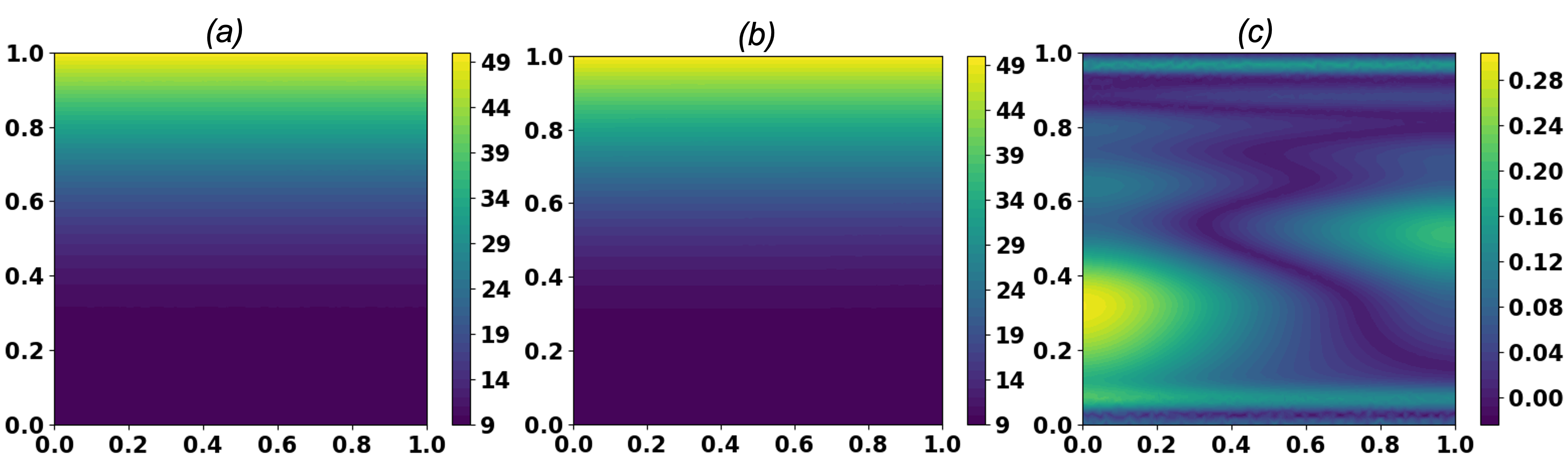}
    \caption{Comparison between the results obtained using the PI-TCN model and the fully coupled finite element method at $t=990$s. The PI-TCN model is trained using the coarse mesh, $N_{elem}=702$, and provides the nodal $\Theta$ for the fine mesh, $N_{elem}=7722$. a) $\Theta$ (in °C) from the PI-TCN model, b) $\Theta$ (in °C) using the FEM, c) $Er_{rel}^{\boldsymbol{x}}(\Theta)$ (in $\%$).} 
    \label{plate_50mesh_rel_error}
\end{figure}

Like the previous example, the PI-TCN model is trained using the boundary data and Gauss points. Then, the nodal points are used to check the model's validity after the training is complete, i.e., the nodal points are not used throughout the training process. Figures \ref{2d_t39} and \ref{2d_t49} are based on nodal points. This indicates that the PI-TCN model generalizes to unseen data. Now, we investigate how the trained PI-TCN model generalizes to meshes different from the relatively coarse mesh used for training. Figure \ref{plate_50mesh_rel_error} compares $\Theta$ for a relatively fine mesh ($N_{elem}=7722$) obtained from the PI-TCN model trained on a coarse mesh ($N_{elem}=702$) and $\Theta$ attained from the fully coupled finite element analysis. The PI-TCN accurately predicts the temperature variation (with a maximum $Er_{rel}^{\boldsymbol{x}}(\Theta)$ of around $0.3\%$), although training on a coarse mesh.

\begin{figure}[!htb]
    \centering
\includegraphics[width=1.0\textwidth]{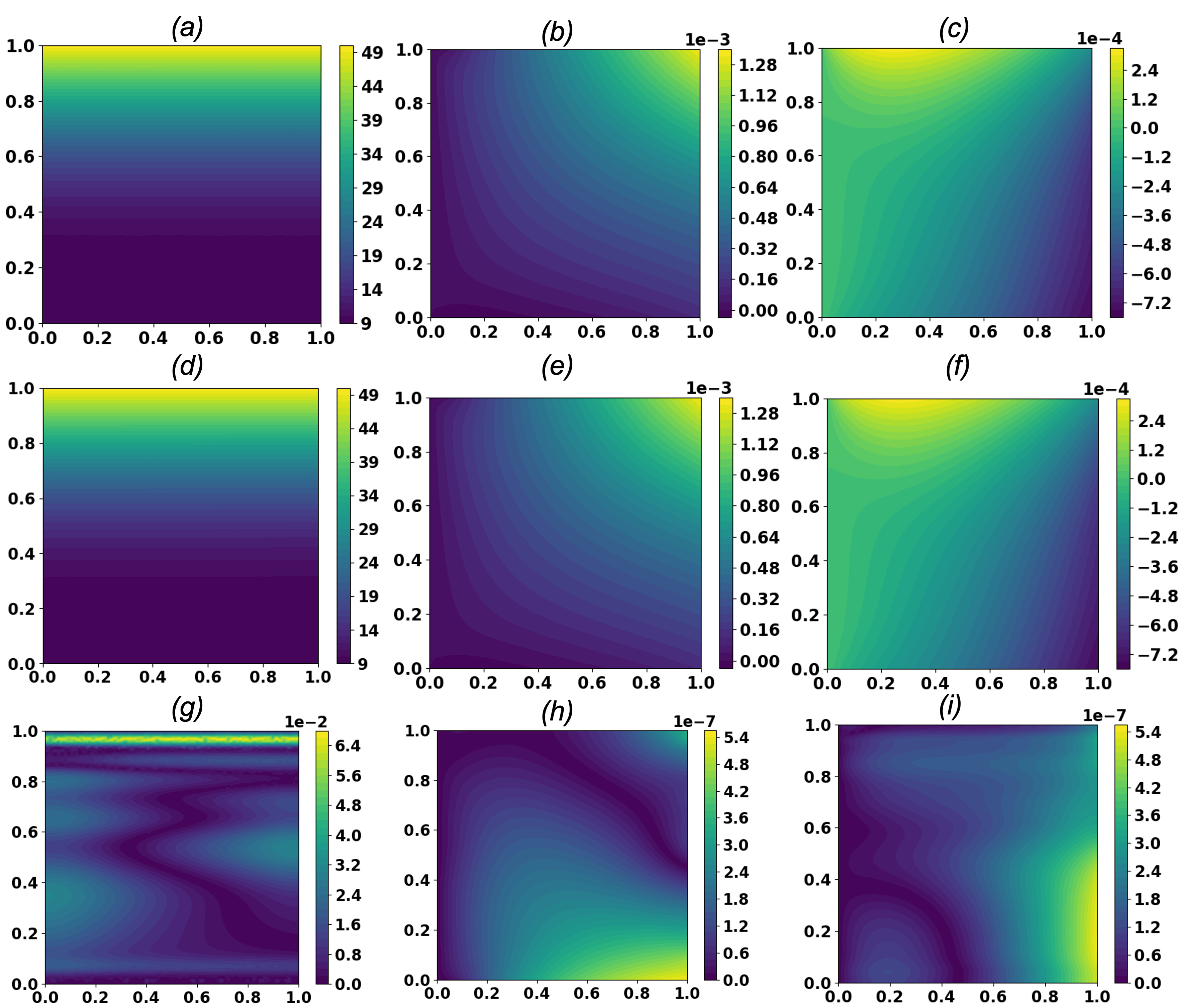}
    \caption{Comparison between the results obtained using the I-FENN framework and finite element method at $t=990$s, when the PI-TCN model is trained using the coarse mesh, $N_{elem}=702$, and provides the nodal $\Theta$ for the fine mesh, $N_{elem}=7722$. Then, the trained PI-TCN model is integrated inside the mechanical weak form to obtain the displacements for the fine mesh. Temperatures are in °C, and displacements are in m. a) $\Theta$ from the I-FENN framework, b) $u$ from the I-FENN framework, c) $v$ from the I-FENN framework, d) $\Theta$ using the FEM, e) $u$ using the FEM, f) $v$ using the FEM, g) $Er_{abs}^{\boldsymbol{x}}(\Theta)$, h) $Er_{abs}^{\boldsymbol{x}}(u)$, and i) $Er_{abs}^{\boldsymbol{x}}(v)$.} 
    \label{2dplate_mesh50}
\end{figure}

\begin{figure}[!htb]
    \centering
\includegraphics[width=0.8\textwidth]{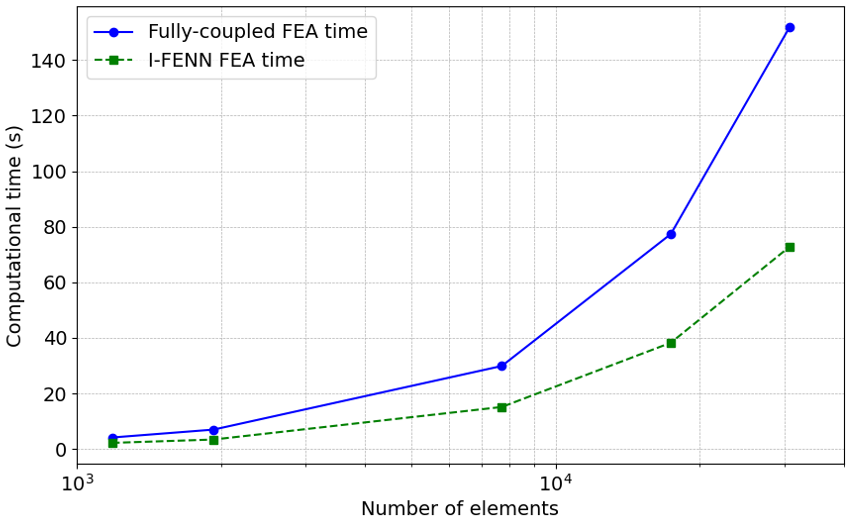}
    \caption{Computational time comparison between the fully coupled finite element analysis and the I-FENN finite element analysis.} 
    \label{2d_comp_time}
\end{figure}

After training the PI-TCN model trained using the coarse mesh, the PI-TCN model is integrated inside the finite element scheme to solve the mechanical weak form for the fine mesh ($N_{elem}=7722$). Figure \ref{2dplate_mesh50} displays the predictions obtained using the I-FENN framework for the fine mesh, where the PI-TCN model is trained using the coarse mesh. Figure \ref{2dplate_mesh50} implies that the PI-TCN model generalizes across different meshes. Figure \ref{2d_comp_time} illustrates the computational time required for the monolithic and I-FENN finite element analyses (not accounting for training time) using different mesh sizes. The PI-TCN needs to be trained only once, and then, it can be used across different problems. More details about the computational saving are available in Section \ref{1d_comput_time}.

\subsection{Plate with a hole}\label{Plate_hole2D}

\begin{figure}[!htb]
    \centering
\includegraphics[width=0.5\textwidth]{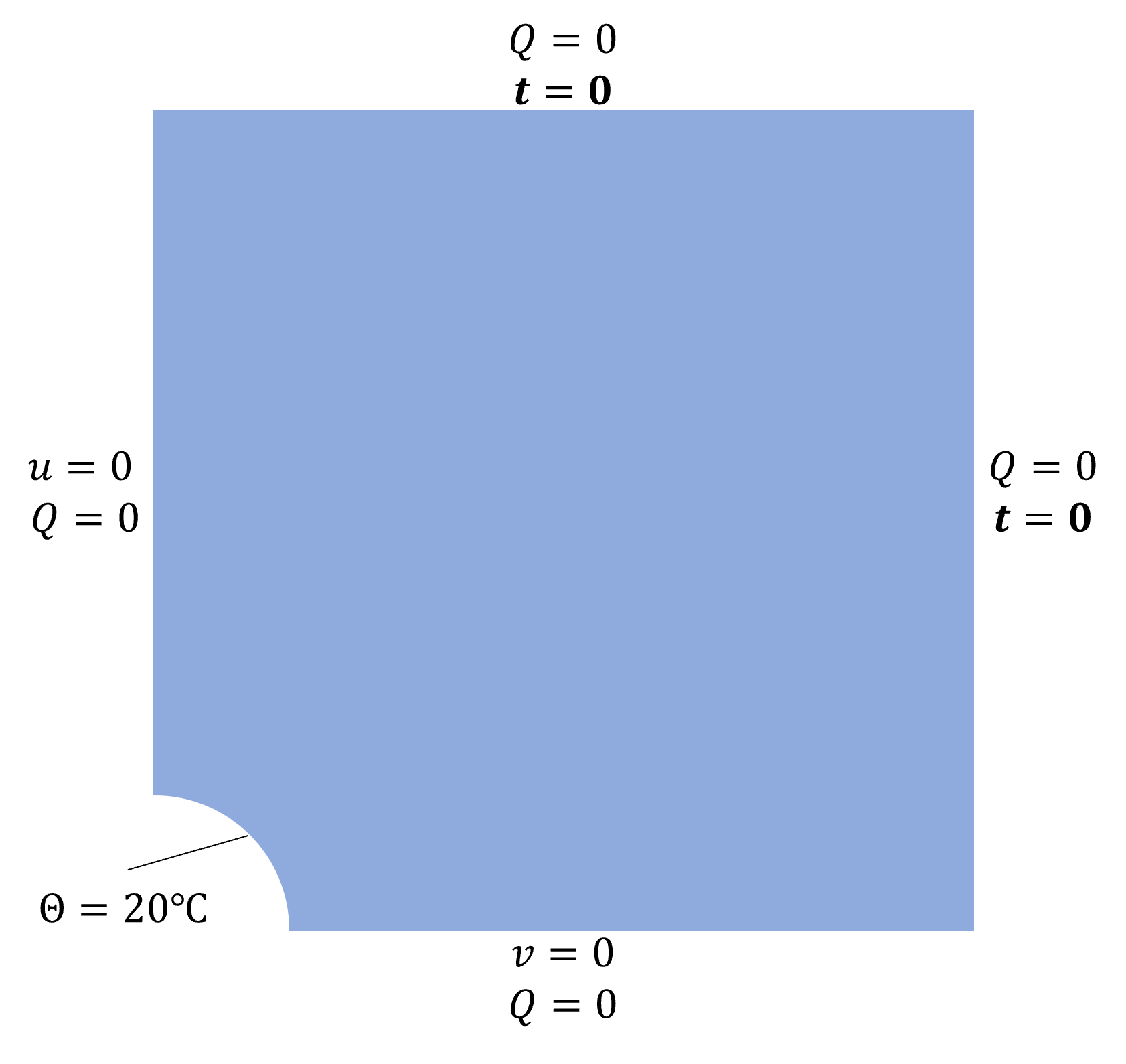}
    \caption{Schematic illustration of the geometry and boundary conditions of the 2D plate with a hole problem.}
    \label{plate_hole}
\end{figure}

\begin{figure}[!htb]
    \centering
\includegraphics[width=1.0\textwidth]{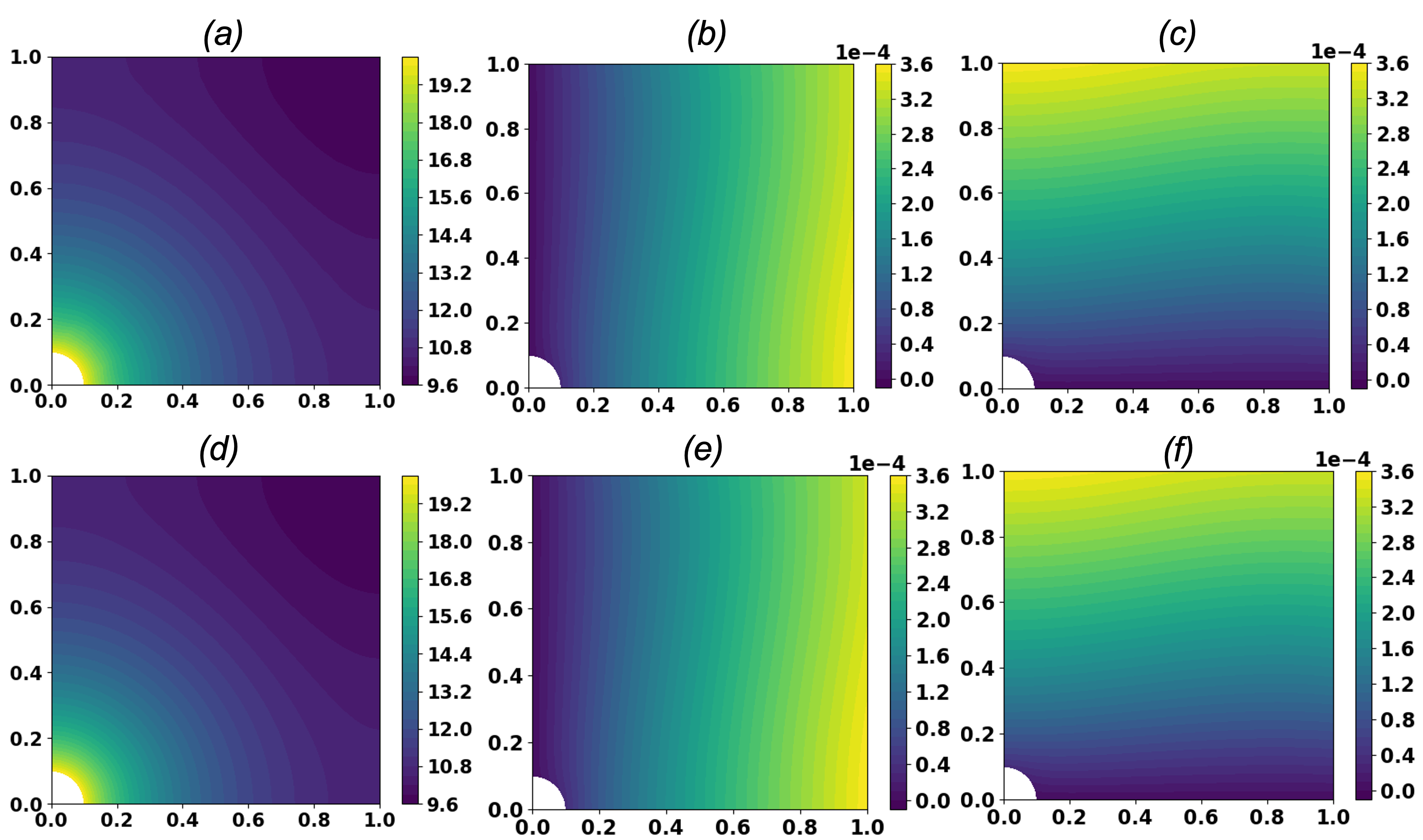}
    \caption{Finite element solutions for two different mesh sizes: a) $\Theta$ for $N_{elem}=803$, b) $u$ for $N_{elem}=803$, c) $v$ for $N_{elem}=803$, d) $\Theta$ for $N_{elem}=7819$, e) $u$ for $N_{elem}=7819$, and f) $v$ for $N_{elem}=7819$.}
    \label{2D_plate_hole_mesh}
\end{figure}

Another example we consider is a plate with a hole at the center with a prescribed temperature on the hole boundary. Due to loading and geometric symmetries, a quarter of the plate is considered, and symmetric boundary conditions are applied on the symmetry planes. Zero tractions and fluxes boundary conditions are imposed on the plate's outer boundary. The boundary conditions are illustrated in Figure \ref{plate_hole}. The quarter plate has a side of unit length, and the radius of the hole is $R=0.1$m. A finite element model is solved using the fully coupled analysis to generate the training data. The finite element analysis uses 50-time increments that are spaced logarithmically. A mesh dependence study is performed, as shown in Figure \ref{2D_plate_hole_mesh}. When a mesh that is ten times finer is used, no significant deviations are noticed.

\begin{figure}[!htb]
    \centering
\includegraphics[width=0.4\textwidth]{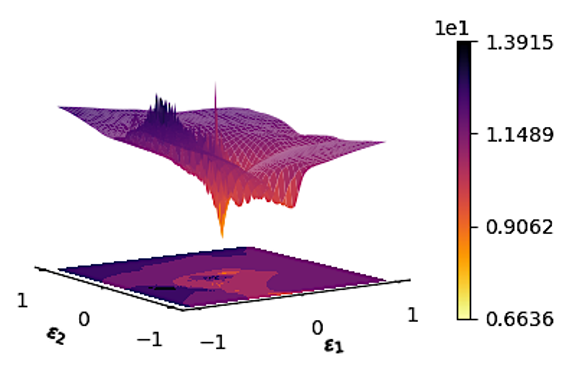}
    \caption{Loss landscapes of the PI-TCN model for the 2D plate with a hole example.}
    \label{2dhole_losslandscape}
\end{figure}

\begin{figure}[!htb]
    \centering
\includegraphics[width=1.0\textwidth]{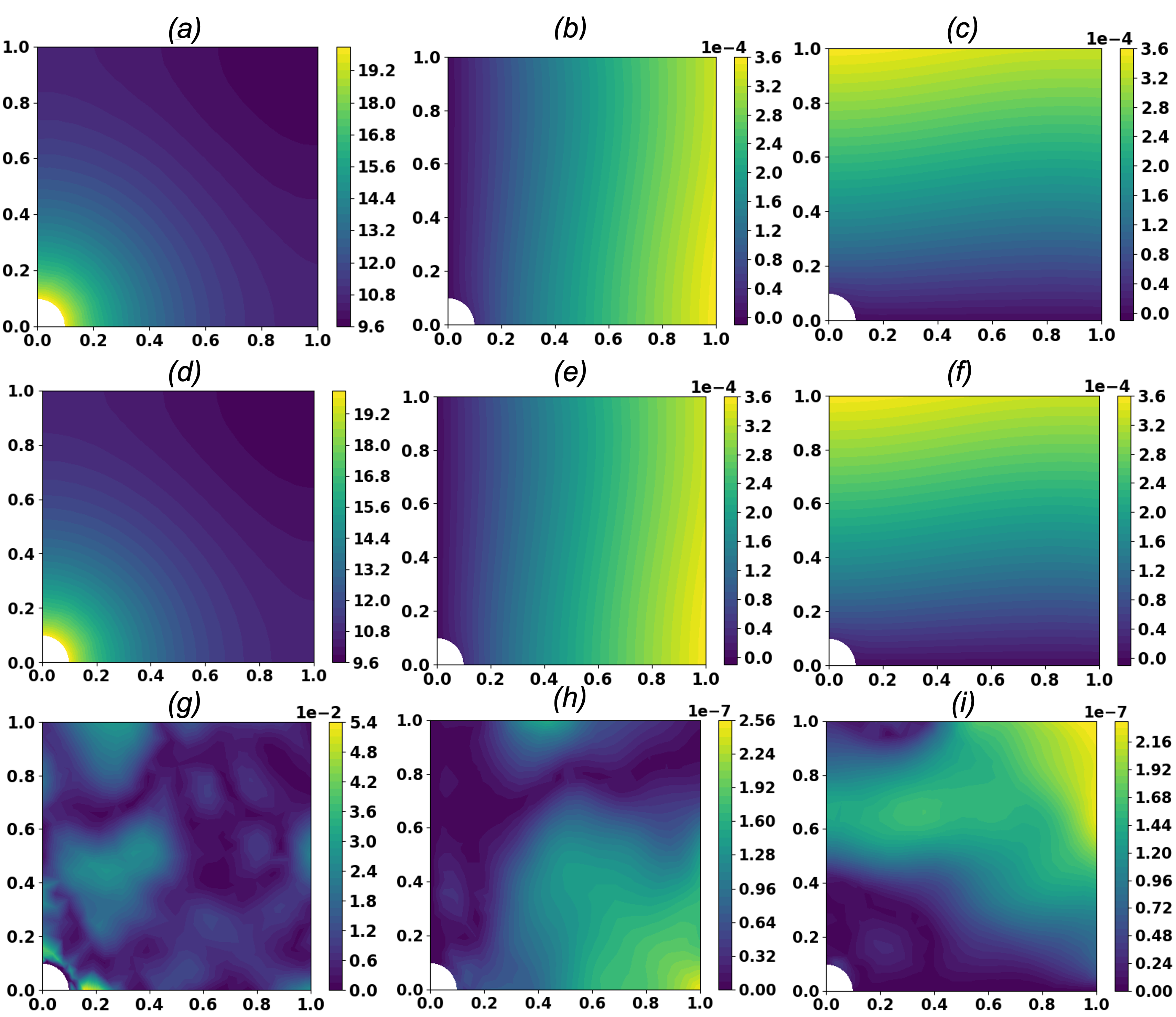}
    \caption{Comparison between the results obtained using the I-FENN framework and finite element method at $t=990$s. Temperatures are in °C, and displacements are in m. a) $\Theta$ from the I-FENN framework, b) $u$ from the I-FENN framework, c) $v$ from the I-FENN framework, d) $\Theta$ using the FEM, e) $u$ using the FEM, f) $v$ using the FEM, g) $Er_{abs}^{\boldsymbol{x}}(\Theta)$, h) $Er_{abs}^{\boldsymbol{x}}(u)$, and i) $Er_{abs}^{\boldsymbol{x}}(v)$.} 
    \label{2d_hole_results}
\end{figure}

After that, the PI-TCN model is trained to capture the energy equation and infer the temperature variation $\Theta$ based on the time increments, coordinates, and strain rate (see Figure \ref{tcn_I-FENN}). A mesh size of $N_{elem}=803$ is utilized for the training of the PI-TCN model. Next, we visualize the loss landscapes, as depicted in Figure \ref{2dhole_losslandscape}. The two-dimensional loss landscapes exhibit a rough and uneven terrain with a distinct minimum the model converges to when trained due to the incorporation of the curvature information of the loss function into the training process \cite{basir2023investigating}. Then, the trained PI-TCN model is integrated into the finite element analysis (see Figure \ref{I-FENN_fig}). The PI-TCN model is used during the finite element analysis to predict the temperature variation $\Theta$. Then, $\Theta$ is used within the weak form of the balance of linear momentum equation to determine the displacements. To calculate errors, we use the solutions obtained using the finite element model that is fully coupled as the reference. Figure \ref{2d_hole_results} compares the results obtained from the fully coupled finite element analysis and the I-FENN framework, indicating that both agree.

\begin{figure}[!htb]
    \centering
\includegraphics[width=1.0\textwidth]{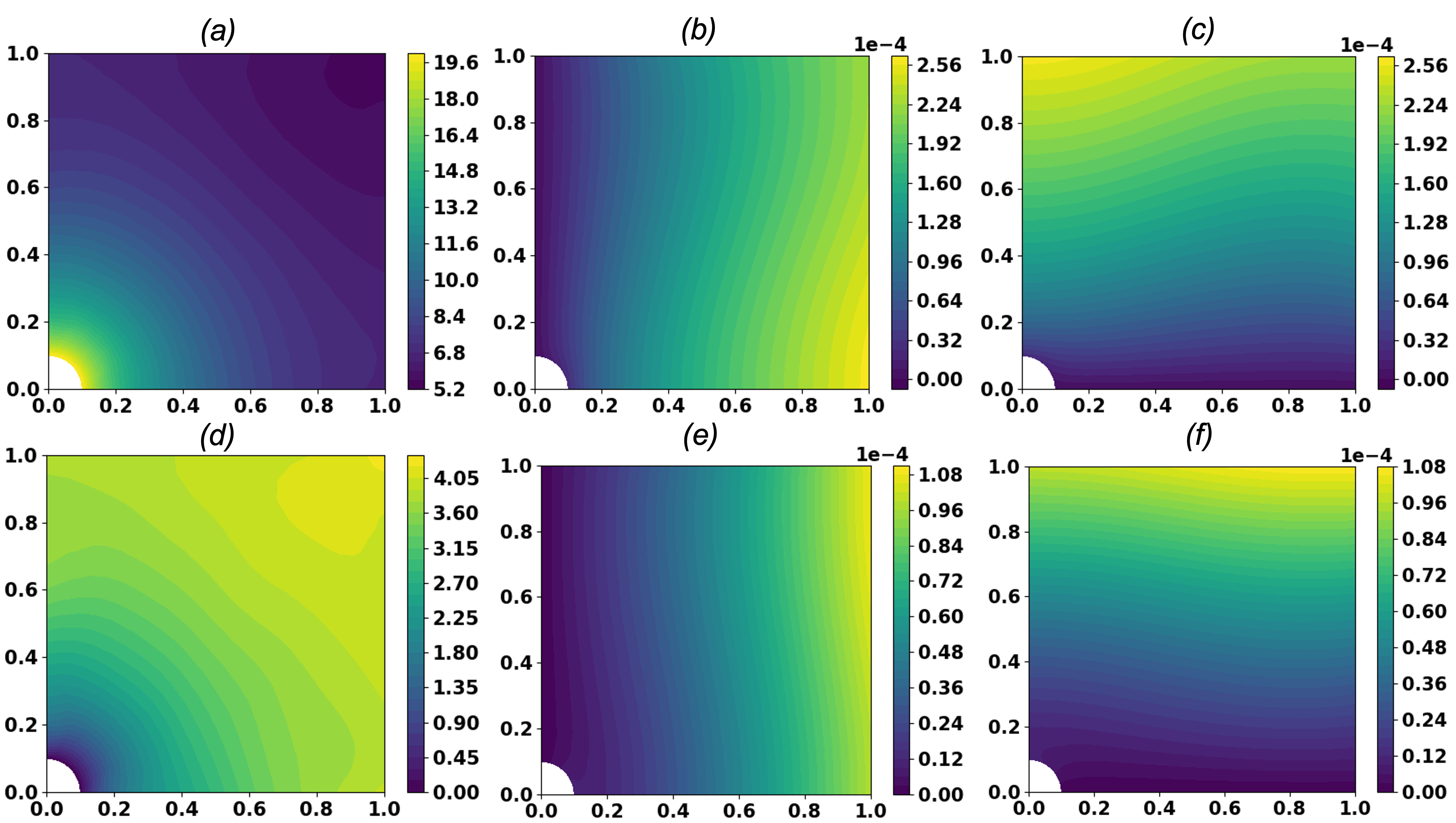}
    \caption{I-FENN solution using a PINN model with 1200 trainable parameters. The solution reported here is at $t=990$s. Temperatures are in °C, and displacements are in m. a) temperature variation $\Theta$, b) displacement along the $x-$axis $u$, c) displacement along the $y-$axis $v$, d) $Er_{abs}^{\boldsymbol{x}}(\Theta)$, e) $Er_{abs}^{\boldsymbol{x}}(u)$, and f) $Er_{abs}^{\boldsymbol{x}}(v)$.} 
    \label{mlp_smallNet_results}
\end{figure}

\begin{figure}[!htb]
    \centering
\includegraphics[width=1.0\textwidth]{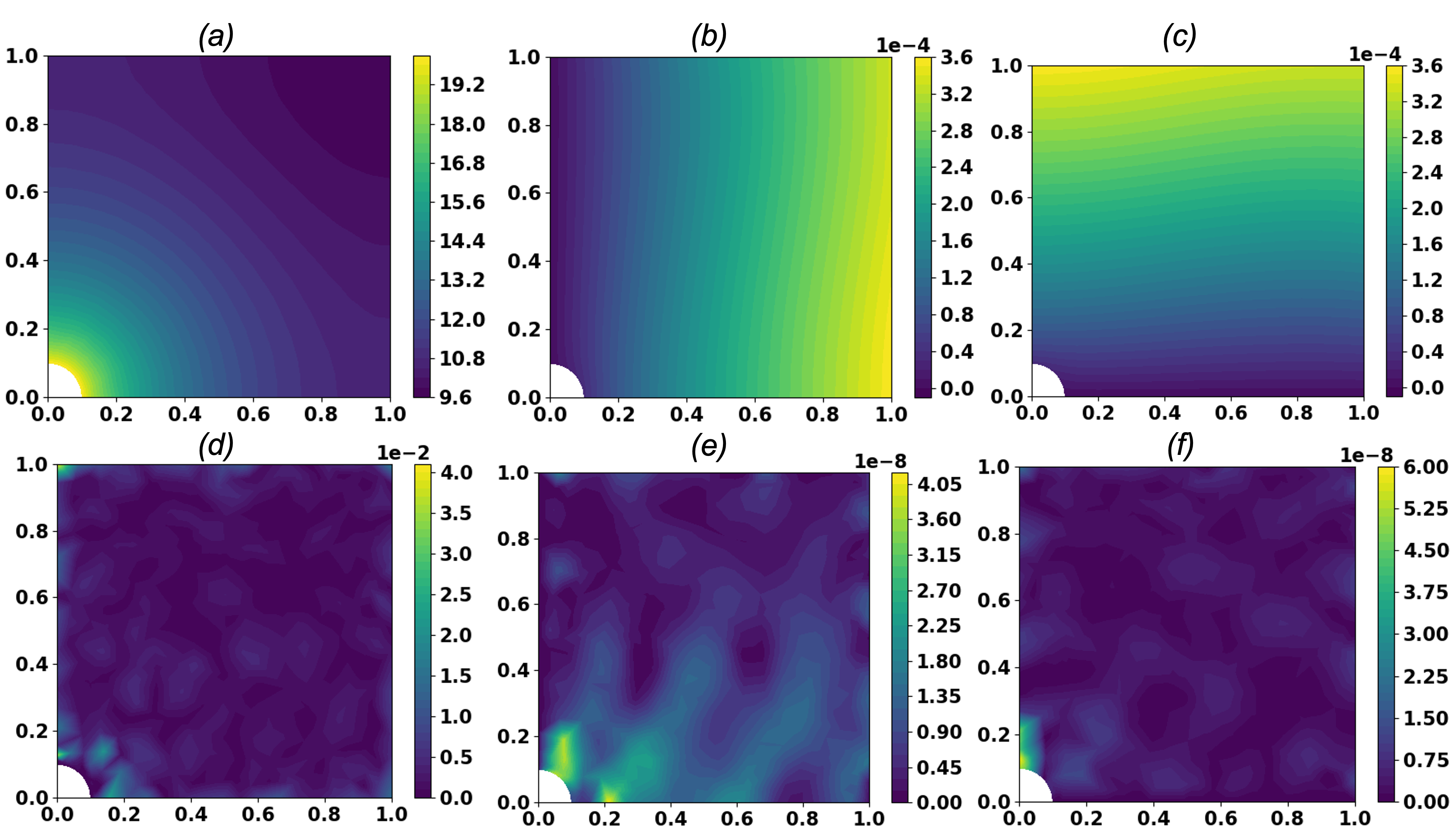}
    \caption{I-FENN solution using a PINN model with 12k trainable parameters. The solution reported here is at $t=990$s. Temperatures are in °C, and displacements are in m. a) temperature variation $\Theta$, b) displacement along the $x-$axis $u$, c) displacement along the $y-$axis $v$, d) $Er_{abs}^{\boldsymbol{x}}(\Theta)$, e) $Er_{abs}^{\boldsymbol{x}}(u)$, and f) $Er_{abs}^{\boldsymbol{x}}(v)$.} 
    \label{mlp_largeNet_results}
\end{figure}

\begin{figure}[!htb]
    \centering
\includegraphics[width=1.0\textwidth]{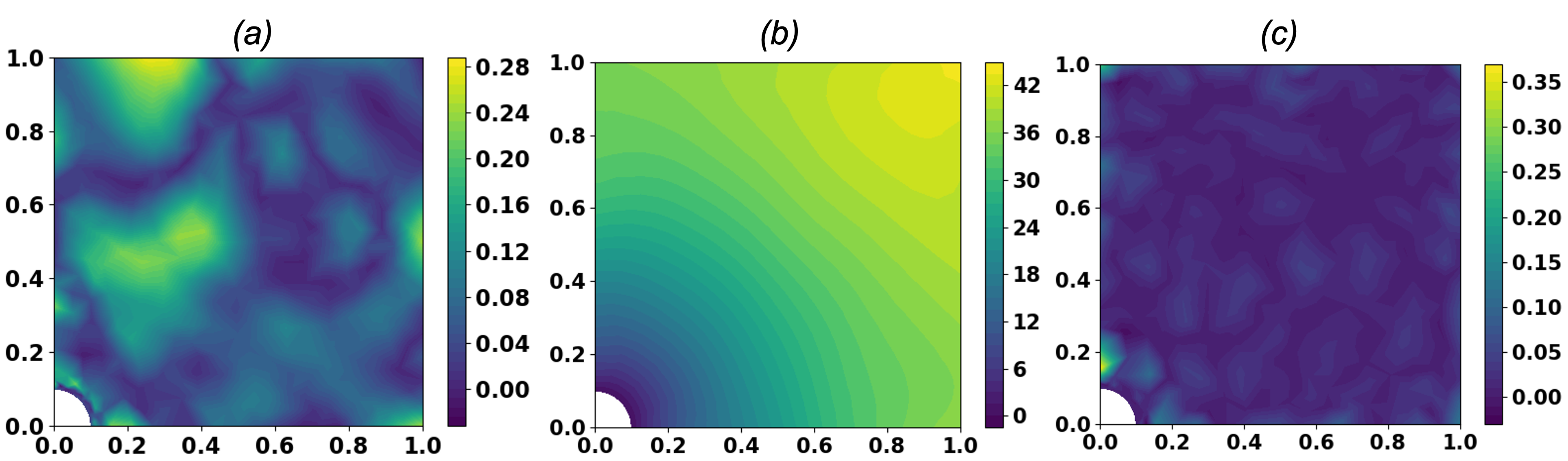}
    \caption{Relative error comparison for the temperature variation $\Theta$ at $t=990$s obtained using the a) PI-TCN model (around 44k trainable parameters), b) relatively small-sized PINN model (around $1200\times50=60\text{k}$ trainable parameters), and relatively large-sized PINN model (around $12\text{k}\times50=600\text{k}$ trainable parameters). The values of the relative errors are in $\%$.} 
    \label{pitcn_vs_mlp_rel_comparison}
\end{figure}

Here, we focus on comparing the performance of the PINN and PI-TCN models. We solve the problem presented in Figure \ref{plate_hole} using the I-FENN framework (see Figure \ref{I-FENN_fig}), and we consider PI-TCN and PINN models. When developing a PINN-based I-FENN framework, we train a PINN model for each time increment. In order to make a fair comparison between the PINN and PI-TCN models, it is necessary to consider models of similar size. The PI-TCN has around 44k trainable parameters for all 50-time increments. Since each time increment has a PINN model, let us consider a PINN model with 1200 trainable parameters, where the same architecture is used for each time increment, but each one has its optimized weights and biases. Figure \ref{mlp_smallNet_results} shows the solution obtained using an I-FENN framework with a similar number of trainable parameters as the PI-TCN. Considering Figures \ref{2d_hole_results} and \ref{mlp_smallNet_results}, it is clear that the PI-TCN model outperforms the PINN model. The errors are calculated using the fully coupled finite element analysis as a reference. To show that PINNs can also be used within the I-FENN framework, we solve the problem again but with a larger network. Let us consider a PINN model with around 12k trainable parameters, i.e., the total number of trainable parameters is around $50\times12\text{k} = 600\text{k}$. Figure \ref{mlp_largeNet_results} depicts the results for such a network. The results using such a network are in good agreement with those obtained from the fully coupled finite element analysis, as shown in the error plots. We use transfer learning for the training of both PINN models. In other words, the weights and biases of the PINN model at a time step $t_n$ are used as initial guesses for the weights and biases of the PINN model at time step $t_{n+1}$. Otherwise, the training will be even more challenging for both PINN models discussed above. Transfer learning for the PI-TCN model is not applicable here as the model is simultaneously trained for all time steps. Figure \ref{pitcn_vs_mlp_rel_comparison} summarizes the performance of the PI-TCN model and the two PINN models. The maximum relative error for the PI-TCN model is around 0.28\%, and this error is not concentrated in a specific region in the domain. The relatively small-sized PINN model fails to capture the response. The relatively large-sized PINN model has a maximum relative error of around 0.35\% near the hole boundary. Notably, this PINN model has a very low relative error in regions far from the hole boundary. 

\subsection{3D Plate}\label{3DPlate}

\begin{figure}[!htb]
    \centering
    \includegraphics[width=0.85\textwidth]{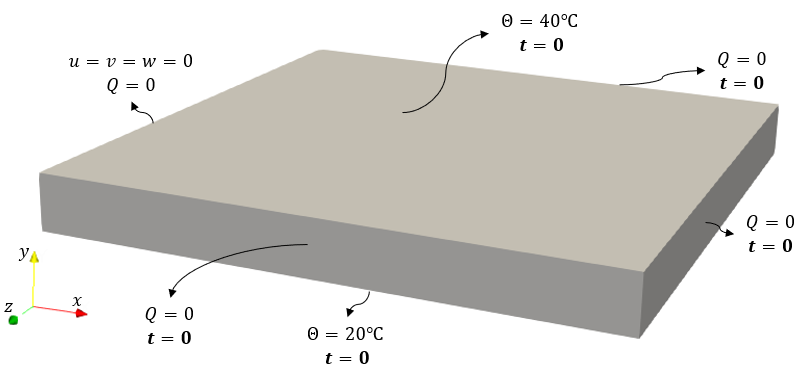}
    \caption{Schematic illustration of the geometry and boundary conditions of the 3D plate problem.}
    \label{3d_plate_geom_bc}
\end{figure}

The last numerical example we consider is a 3D plate; Figure \ref{3d_plate_geom_bc} shows the geometry and boundary conditions considered for this example. Temperature is prescribed at the top ($+y$-axis) and bottom ($-y$-axis) faces. The plate is clamped at the $-x$-axis face. Flux conditions $Q=\boldsymbol{q}\cdot\boldsymbol{n}=0$ are applied on the $+x$-axis, $-x$-axis, $+z$-axis, and $-z$-axis faces of the plate. Additionally, zero tractions $\boldsymbol{t}=\boldsymbol{0}$ are applied on the $+x$-axis, $+y$-axis, $-y$-axis, $+z$-axis, and $-z$-axis faces of the plate. The dimensions of the plate are  $1.0 \text{m} \times 0.1 \text{m} \times 1.0 \text{m}$. A fully coupled finite element model based on the monolithic approach with 2400 tetrahedral elements is initially constructed for the purpose of producing training data. In the finite element analysis, a total of fifty-time increments, spaced logarithmically, are employed.

\begin{figure}[!htb]
    \centering
\includegraphics[width=1.0\textwidth]{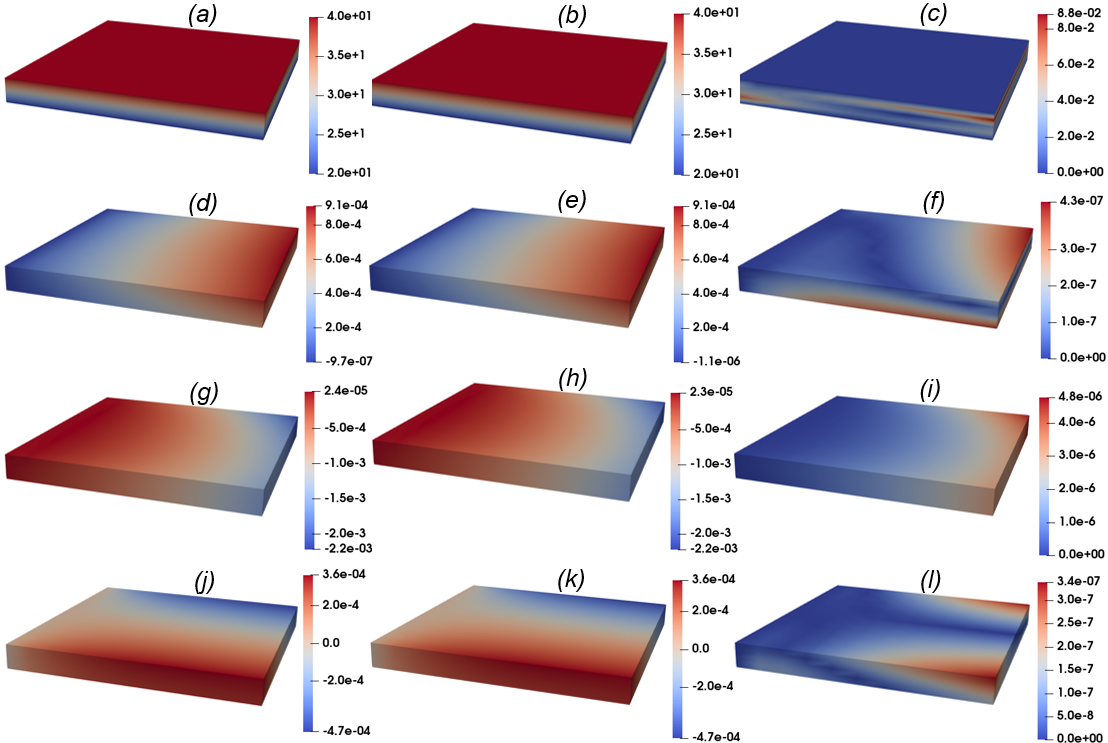}
    \caption{Comparison between the results obtained using the I-FENN framework method at $t=990$s. Temperatures are in °C, and displacements are in m. a) $\Theta$ from the I-FENN framework, b) $\Theta$ using the FEM, c) $Er_{abs}^{\boldsymbol{x}}(\Theta)$, d) $u$ from the I-FENN framework, e) $u$ using the FEM, f) $Er_{abs}^{\boldsymbol{x}}(u)$, g) $v$ from the I-FENN framework, h) $v$ using the FEM, i) $Er_{abs}^{\boldsymbol{x}}(v)$. j) $w$ from the I-FENN framework, k) $w$ using the FEM, and l) $Er_{abs}^{\boldsymbol{x}}(w)$.} 
    \label{3d_plate_results}
\end{figure}

\begin{figure}[!htb]
    \centering
\includegraphics[width=0.8\textwidth]{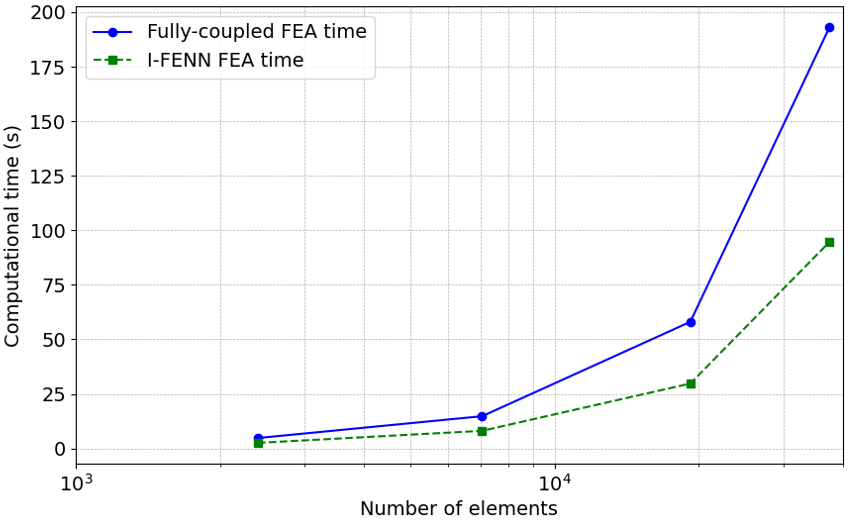}
    \caption{Computational time comparison between the fully coupled finite element analysis and the I-FENN finite element analysis.} 
    \label{3dplate_comp_time}
\end{figure}

Subsequently, the PI-TCN model undergoes training to learn the energy equation and predict the temperature variations $\Theta$ contingent upon the time, spatial coordinates, and strain rate. The PI-TCN model is trained using boundary data and Gauss points. The model's validity is assessed using nodal points, which are not incorporated during the training phase. The trained PI-TCN model is incorporated into the finite element analysis, as illustrated in Figure \ref{I-FENN_fig}. During the finite element analysis, the PI-TCN model enables the prediction of temperature variation $\Theta$, which is utilized in the weak formulation of the linear momentum balance equation to find the displacements. Figure \ref{3d_plate_results} depicts the temperature and displacement profiles obtained from the I-FENN framework and the fully coupled finite element analysis. Besides, Figure \ref{3d_plate_results} shows the error for the different profiles, where the fully coupled FEA is used as the reference solution. It is observed that the results attained from the I-FENN framework are in agreement with those from the fully coupled FEA. Figure \ref{3dplate_comp_time} presents the time required by both the monolithic and I-FENN finite element methods, excluding the duration for training, across various mesh sizes. The PI-TCN, once trained, is applicable to diverse problems without the need for retraining. For further elaboration on the reductions in computational effort, refer to Section \ref{Section_ifenn_overview} and Section \ref{1d_comput_time}.

\section{Discussion, conclusions, and future directions}\label{conclu}

Physics-informed machine learning techniques are gaining increasing attention in the research community to solve partial differential equations (PDEs). This paper proposes an integrated finite element neural network (I-FENN) framework to speed up finite element computations by leveraging deep learning in thermoelasticity. This framework incorporates pre-trained physics-informed machine learning models into the finite element scheme, allowing them to work with element interpolation functions to calculate element-level variables (e.g., strains and stresses), where the trained network calculates the temperatures required for thermoelasticity formulation. We discuss two physics-informed machine learning models: a conventional PINN model (based on MLP) and a physics-informed temporal convolutional network (PI-TCN). One disadvantage of the PINN model is that one needs to develop a PINN model for every time increment required to solve the problem. The PI-TCN is based on convolutional layers, where filters to the input's local regions effectively detect and share local features or patterns. To the best of our knowledge, this is the first implementation of TCN in developing a physics-informed machine learning model. The I-FENN framework, which is based on a PI-TCN model, is used to solve three numerical examples. We show that the PI-TCN model outperforms the data-driven TCN and highlight the I-FENN's gains in computational time. Furthermore, we show that the I-FENN framework is a promising tool, as it opens the door for scaling the transient thermoelasticity problem discussed here and other multiphysics problems. For the transient thermoelasticity problem, after the training of the PI-TCN model, the energy equation is computationally decoupled from the mechanical weak form, while the temperature effect can be inferred from the PI-TCN model for any fine mesh, although being trained using a relatively coarse mesh. Additionally, we also conducted a comparison between the conventional PINN and PI-TCN models, and our findings demonstrate that the PI-TCN model performs better than the PINN model.

Although physics-informed machine learning algorithms have been proven successful in solving PDEs governing various physical phenomena, most of this work has been confined to academic settings because implementing physics-informed machine learning for real-world purposes is challenging and demands further understanding. This paper does not claim to provide the final words on how machine learning algorithms can solve mechanics problems more effectively. There are numerous challenges associated with utilizing machine learning to solve mechanics problems. For instance, the optimization problem required for determining the weights and biases of a machine learning model is often nonconvex, increasing the possibility of becoming trapped in local minima, as the loss landscape plots for the different examples have revealed. Moreover, the network architecture and hyperparameters were chosen through a trial-and-error process in this study. However, a more robust technique for selecting architecture and hyperparameters that yield high accuracy is required. Pantidis et al. \cite{PantidisError2023} showed that the network's topology of a PINN model plays a vital role in controlling its performance. They observed performance patterns with the network's topology for the developed PINN model. Additionally, Hamdia et al. \cite{hamdia2021efficient}, Chadha et al. \cite{chadha2022optimizing}, and Wang et al. \cite{wang2022auto} suggested employing optimization algorithms to determine the architecture and hyperparameters for machine learning models systematically. 

Another challenge when employing physics-informed machine learning for solving PDEs is the instability associated with automatic differentiation (AD) for calculating gradients. The stability of AD-based gradient calculation is beyond the scope of this paper. Nonetheless, additional information regarding the stability of automatic differentiation within the physics-informed machine learning models and potential alternatives to address the instability in gradient computation can be found in the work of Basir \cite{basir2023investigating} and He et al. \cite{he2023use}. AD stability becomes more pertinent in regions with higher solution gradients, such as boundaries. Another challenge when developing a physics-informed machine learning model is the imposition of boundary conditions. Although, in this paper, we used distance functions to enforce the Dirichlet boundary conditions strictly, the Neumann boundary conditions are applied through weak enforcement. Specifically, the Neumann boundary conditions are imposed with a penalty term added to the definition of the loss function. The relatively higher error values in the boundary region in the discussed examples could be attributed to the abovementioned points. It is worth mentioning that even coming up with analytical distance functions to apply the Dirichlet boundary conditions, as we do in the current work, can be inadmissible for irregular geometries or even different variations of boundary conditions used on relatively simple geometries. The imposition of boundary conditions in the context of physics-informed machine learning models is an active area of research \cite{rao2021physics, sukumar2022exact, basir2022physics}.
\section*{Acknowledgment}
The authors wish to express their gratitude to the NYUAD Center for Research Computing for their provision of resources, services, and skilled personnel. They would like to give special recognition to Fatema Salem Alhajeri for her help and support throughout this project.
\section*{Data and code availability}
The data and source code that support the findings of this study can be found at \url{https://github.com/DiabAbu/PI-TCN_I-FENN_Thermoelasticity}. \textcolor{red}{Note to editor and reviewers: the link above will be made public upon the publication of this manuscript. The data and source code can be made available upon request during the review period.}

\bibliography{mybibfile}

\end{document}